\documentclass[nofootinbib,twocolumn, superscriptaddress]{revtex4-2}
\usepackage[utf8]{inputenc}

\usepackage{xcolor}
\usepackage{amsmath,systeme}
\usepackage{amsfonts}
\usepackage{natbib}
\usepackage{graphicx}
\usepackage{lipsum}
\usepackage{layout}
\usepackage{cancel}
\usepackage{wasysym}
\usepackage{booktabs}
\usepackage{comment}
\usepackage[normalem]{ulem}
\usepackage{hyperref}
\hypersetup{
    colorlinks,
    linkcolor={blue!80!black},
    citecolor={blue!80!black},
    urlcolor={blue!80!black}
}
\hypersetup{hidelinks}
\usepackage{xifthen}
\usepackage{fontawesome}

\newcommand{\Rxx}{C_\ell}

\newcommand\pdfref[3]{%
    \href{phd://open-paper?id=#1&page=#2}{%
    \textup{[\textbf{\ifthenelse{\isempty{#3}}{here}{#3}}]}}%
}

\newcommand{\ds}{1}
\newcommand{\dss}{2}

\newcommand{\p}{_{p_1}}
\newcommand{\pp}{_{p_2}}
\newcommand{\q}{_{q_1}}
\newcommand{\qq}{_{q_2}}

\begin{document}

\title{CMB component-separated power spectrum estimation by Spectral Internal Linear Combination (SpILC)}

\author{Jack Y.\ L.\ Kwok}
\email[E-mail:\ ]{ylk26@cam.ac.uk}
\affiliation{DAMTP, Centre of Mathematical Sciences, University of Cambridge, Wilberforce Road, Cambridge CB3 0WA, United Kingdom}
\affiliation{Kavli Institute for Cosmology Cambridge, Madingley Road, Cambridge CB3 0HA, UK}
\author{William R.\ Coulton}
\affiliation{Kavli Institute for Cosmology Cambridge, Madingley Road, Cambridge CB3 0HA, UK}
\affiliation{DAMTP, Centre of Mathematical Sciences, University of Cambridge, Wilberforce Road, Cambridge CB3 0WA, United Kingdom}
\author{Niall MacCrann}
\affiliation{DAMTP, Centre of Mathematical Sciences, University of Cambridge, Wilberforce Road, Cambridge CB3 0WA, United Kingdom}
\affiliation{Kavli Institute for Cosmology Cambridge, Madingley Road, Cambridge CB3 0HA, UK}
\author{Fiona McCarthy}
\affiliation{DAMTP, Centre of Mathematical Sciences, University of Cambridge, Wilberforce Road, Cambridge CB3 0WA, United Kingdom}
\affiliation{Kavli Institute for Cosmology Cambridge, Madingley Road, Cambridge CB3 0HA, UK}
\affiliation{Center for Computational Astrophysics, Flatiron Institute, 162 5th Avenue, New York, NY 10010 USA}
\author{Boris Bolliet}
\affiliation{Astrophysics Group, Cavendish Laboratory, J.\ J.\ Thomson Avenue, Cambridge CB3 0HE, United Kingdom}
\affiliation{Kavli Institute for Cosmology Cambridge, Madingley Road, Cambridge CB3 0HA, UK}
\author{Blake D.\ Sherwin}
\affiliation{DAMTP, Centre of Mathematical Sciences, University of Cambridge, Wilberforce Road, Cambridge CB3 0WA, United Kingdom}
\affiliation{Kavli Institute for Cosmology Cambridge, Madingley Road, Cambridge CB3 0HA, UK}
\date{\today}

\begin{abstract}
	Component separation methods mitigate the cross-contamination between different extragalactic and galactic contributions to cosmic microwave background (CMB) data. This is often done by linearly combining CMB \emph{maps} from different frequency channels using internal linear combination (ILC) methods. We demonstrate that deriving power spectrum estimators directly by linearly combining auto- and cross-spectra instead of maps allows us to obtain a different constrained-optimization problem that allows fewer (deprojection) constraint equations than combining at map level using the constrained ILC method. Through simulations, we show that our Spectral internal linear combination (SpILC) produces CMB power spectrum estimators with more than 7 times smaller errorbars than constrained ILC (with thermal Sunyaev-Zel'dovich and cosmic infrared background deprojections) at $\ell\gtrsim 4000$ for Simons Observatory-like observations. Spectral ILC outperforms constrained ILC methods when some modeled components are spatially uncorrelated, e.g.\ the primary CMB is uncorrelated with foregrounds, and the difference in performance is most significant at noise-dominated scales. 
    More generally, our work shows that component-separated maps with foreground deprojections do not necessarily produce minimum-variance two-or-higher-point estimators.
\end{abstract}

\maketitle

\section{Introduction}

Over the past decade, ground-based cosmic microwave background (CMB) experiments e.g.\ Atacama Cosmology Telescope (ACT) \cite{2015arXiv151002809H,act_dr4_maps} and South Pole Telescope (SPT) \cite{spt_carlstrom,spt_release1,2014SPIE.9153E..1PB} have built upon space-based observations from \textit{Planck} \cite{2020A&A...641A...1P}, pushing the resolution of microwave observations to arcminute scales. This opens up the study of CMB \emph{secondary anisotropies}---late-time perturbations to the primary CMB---including weak lensing of CMB temperature and polarization \cite{smith_lensing,polarbear_powerspectrum,planck18_lensing,act_lensing_map,spt_lensing23}, the thermal (tSZ) \cite{sunyaev72,sunyaev80,planck_ymap, ymap_coulton, act_sz_catalog, spt_sz_catalog} and kinetic Sunyaev-Zel'dovich (kSZ) effect \cite{sunyaev72,sunyaev80,first_kSZ,schaan20,spt_ksz21,maccrann_4pksz,spt_4pksz}, and the patchy screening effect \cite{dvorkin09,smith_ferraro,patchy_coulton}. 
At the advent of the ground-based CMB experiment Simons Observatory (SO) \cite{so}, complemented by galaxy surveys including the Vera Rubin Observatory \cite{lsst}, the \textit{Euclid} \cite{euclid} and SPHEREx \cite{spherex} space telescopes, the Dark Energy Survey (DES) \cite{des} and the Dark Energy Spectroscopic Instrument (DESI) \cite{desi}, the millimeter sky will continue to improve our understanding of the primordial universe and the growth of cosmic structures.

In order to study the wealth of signals in intensity and polarization maps across frequency channels, one is motivated to isolate the signals from each other. This includes separating  extragalactic signals (e.g. CMB, tSZ, kSZ, cosmic infrared background, radio point sources) from each other and from galactic (e.g.\ dust, synchrotron) emissions.
Various \emph{component separation} methods are devised for this purpose, whereby the different frequency dependencies (\emph{spectral energy distributions}, or SEDs) and spatial properties of components are exploited to produce a map of the desired component.
These methods can be broadly separated into parametric and ``blind'' methods. Parameteric methods include \texttt{Commander}~\cite{eriksen06,eriksen08,planck15_diffuse}, which samples the joint posterior distribution of spectral parameters and component maps using a Bayesian approach. ``Blind'' methods are model agnostic except for the component of interest. Examples are\ \texttt{SEVEM} \cite{sevem03,sevem08,sevem12}, which subtracts off a linear combination of internally-constructed foreground templates from a map to form a minimum-variance CMB map, and Internal Linear Combination (ILC) methods \cite{eriksen_ilc,tegmark_ilc,delabrouille_wmap_2009,pyilc}, which also assume that the signal SED is known, and again seek a minimum-variance component map by a linear combination of maps across frequency channels. The ``semi-blind'' constrained ILC (cILC) \cite{constrained_ilc} method additionally assumes the SEDs of a few modeled foreground components.
{At the power spectrum level, the Analytical Blind Separation (ABS) method \cite{abs1,abs2} performs eigenmode filtering on the observed cross spectra, building upon a unique solution for the signal power in the zero-noise case with more channels than foreground components.}

 
		This paper focuses on the ILC methods (particularly the constrained ILC)---extensively used in \textit{WMAP} \cite{wmap_foreground1,wmap_foreground2}, \textit{Planck} \cite{planck_foreground13,planck_foreground15,planck18_comseparation} and ACT \cite{act_compsep20,ymap_coulton}---which have a minimal set of core assumptions: 1.\ known SEDs for the components to be modeled, 2.\ no spatial correlation between the modeled components and noise (where everything except the modelled signals are called noise), and 3.\ zero \emph{spatial decorrelation}, meaning observed maps across frequency channels have the same underlying component maps.
		For example, consider a data model with two modeled components $s_p$ and  $y_p$, with other uncorrelated components and instrumental noise grouped into the noise term $n_p$. In equation form the data model would look as follows:
\begin{align}
	x^i_p &= a^i s_p + b^i y_p + n^i_p \; ,
	\label{eq:data_model}
\end{align}
where $x^i_p$ is the observed map at frequency channel  $i$, $s_p$ is the component map one wishes to recover, $y_p$ is a modeled component map one wishes to explicitly remove, and  $n^i_p$ are noise maps at channel $i$. The label $p$ denotes pixels, spherical harmonics or needlets for pixel-space, harmonic or needlet ILC, respectively. The SEDs $a^i$ and  $b^i$ are assumed to be known exactly.\footnote{ In the case of the CIB where the SED is not exactly known, the \emph{moment ILC} method \cite{chluba_moment} is found to be an effective mitigation \cite{fiona_cib} (discussion in Sec.\ \ref{sec:6}).}

Under these core assumptions, the cILC method \cite{cilc,chen09} selects a set of weights $w^i$ to linearly combine the observed maps  $x^i_p$ to form minimum-variance \emph{ILC map}  $\hat{s}_p\equiv\sum_i w_i x^i_p$, under the constraints that the maps have unit response to the component of interest,  $\sum_i w_ia^i = 1$, and zero response to all other modeled components,  $\sum_i w_i b^i =0 $; this latter constraint is also called \emph{deprojection}. 
The focus of this study is on extragalactic signals, which in many cases are deprojected rather than minimized together with other sources of noise since residuals may bias further analyses using the component-separated map or spectra \cite{ymap_coulton}, whereas instrumental noise bias can be mitigated using data splits \cite{hill_spergel}. While galactic foregrounds fall at small scales and can be mitigated by masking close to the galactic plane \cite{maccrann23}, their biases after mitigation e.g.\ on lensing measurements for SO need to be carefully studied \cite{irene25}.
 

Our work intends to address the following question: \textbf{what is the unbiased minimum-variance estimator for the foreground-cleaned power spectrum?}
	Here we develop analytical solutions and insights in the case of known SEDs of modeled components. This is an application of the (constrained) ILC formalism to the spectral level, and as such we denote our auto- and cross-spectra estimators as \emph{Spectral ILC} (SpILC) estimators. However, as spectral estimators, SpILC do not recover the full field statistical distribution or higher-order statistics.

While it is true that constrained ILC maps are unbiased minimum-variance estimators \emph{at map level}, are the spectra estimated from the power spectrum of constrained ILC maps---as is typically done in power spectrum estimation in the ILC framework---minimum-variance estimators? Our work demonstrates that this is \emph{not} the case if we can additionally assume that at least two components are spatially uncorrelated, which importantly allows us to obtain a constrained-optimization problem that has fewer (deprojection) constraint equations in the spectral level compared to constrained ILC. This provides a strong case for the constrained Spectral ILC estimators introduced in this paper (Sec.\ \ref{sec:cSpILC}), as we can achieve significantly lower variance than existing constrained ILC methods in scales where noise dominates.

This paper is structured as follows: Sec.\ \ref{sec:2} develops the Spectral ILC formalism, Sec.\ \ref{sec:cSpILC} derives the constrained SpILC estimators---which are to be compared against constrained ILC power spectra for the remainder of the paper; Sec.\ \ref{sec:3} extends them to incorporate data splits; Sec.\ \ref{sec:4} details the simulations used to validate and compare spectra estimated with SpILC and ILC methods; Sec.\ \ref{sec:5} reports results of our validation and comparison; Sec.\ \ref{sec:6} discusses applications of the SpILC estimators.

\section{Spectral ILC (S\lowercase{p}ILC)}
\label{sec:2}
In this paper we introduce the \emph{Spectral ILC} (SpILC), a power spectrum estimator constructed by linearly combining estimated spectra $\hat{C}^{ij}_\ell$ with symmetric weights $W^{ij}_\ell$ (no $\ell$-summation implied): 
\begin{align}
	\hat{K}_\ell \equiv W^{ij}_\ell \hat{C}^{ij}_\ell \; ,
\end{align}
This study specializes to harmonic space throughout. 
For a given component $s$, the \emph{error}  of the ($ss$-spectrum) estimator is defined as
\begin{align}
	\hat{\varepsilon}^{ss}_\ell \equiv \hat{K}^{ss}_\ell - C^{ss}_\ell \; 
\end{align}
where $C^{ss}_\ell$ is the true power spectrum of component $s$.
We say that the SpILC estimator is unbiased if the ensemble average (denoted by angular brackets)
\begin{align}
\langle \hat\varepsilon_\ell  \rangle=0 \; .
\end{align}
Subsequent subsections discuss how the weights $W^{ij}_\ell$ are chosen.

\subsection{Standard Spectral ILC Weights}
This subsection considers the one-component data model
\begin{align}
	x^i_p = a^i s_p + n^i_p \; .
\end{align}
As we specialize to harmonic domain, $p\mapsto (\ell,m)$,  $q\mapsto(\ell',m')$, and  $\langle x^j_p x^{m,*}_q  \rangle =0$ for $p\neq q$. We assume: 1.\ the component SED $a^i$ is known; 2.\ zero spatial signal-noise correlations
\begin{align}
	\langle s_p n^{i,*}_p  \rangle =0 \; ,
\end{align}
for all channels $i$; and 3.\ Gaussianity of the noise $n^i_p$. In this work, we further use the Gaussian approximation on the component(s) $s_p$ to simplify expressions, but as we will show in App.\ \ref{sec:gaussianity} this approximation does \emph{not} enter the calculation of the weights for SpILC.

We impose the unbiased constraint to the spectrum of interest:
\begin{align}
	&W^{ij}_\ell \langle a^i s_p a^j s_p^*  \rangle = \langle s_p s_p^*  \rangle \nonumber \\
	\implies& W_{\ell}^{ij} a^i a^j = 1 \; .
\end{align}
We refer to this as the \emph{normalization constraint}, which ensures that the target component contributes to the recovered power spectrum with the correct amplitude. Subject to this constraint, we optimize the weights $W_{\ell}^{ij}$ such that our figure of merit, the error variance Var($\hat{\varepsilon}_\ell$), is minimized. The constrained-minimization of $\operatorname{Var}(\hat\varepsilon_\ell)$ is equivalent to the constrained-minimization of $\operatorname{Var}(\hat{K}_\ell)$:
\begin{align}
	\frac{\partial}{\partial W_{\ell}^{ij}} \operatorname{Var}(\hat{\varepsilon}_\ell) = \frac{\partial}{\partial W_{\ell}^{ij}}\operatorname{Var}(\hat{K}_\ell-C^{ss}_\ell) = \frac{\partial}{\partial W_{\ell}^{ij}} \operatorname{Var}(\hat{K}_\ell) =0 \; ,
\end{align}
as $C^{ss}_\ell$ is a constant.
The standard SpILC weights are therefore chosen to minimize $\operatorname{Var}(\hat{K}_\ell)$ subjected to $W_{\ell}^{ij}a^i a^j = 1$. We suppose $W_{\ell}^{ij}$ depends on the ensemble-averaged map spectra  $C^{ij}_\ell$ instead of $\hat{C}^{ij}_\ell$, such that it can be taken out of the ensemble average:
 \begin{align}
	 \operatorname{Var}(\hat{K}_\ell) = \operatorname{Var}(W_{\ell}^{jk}\hat{C}^{jk}_\ell) = W_{\ell}^{jk}W_{\ell}^{mn}\operatorname{Cov}(\hat{C}^{jk}_\ell,\hat{C}^{mn}_\ell) \; .
\end{align}
This is a self-consistent assumption, as $W_{\ell}^{ij}$ is chosen to minimize Var($\hat{K}_\ell$), which now depends only on $C^{ij}_\ell$ (as we show immediately below).
The expression for the variance is
\begin{align}
	\operatorname{Var}(\hat{K}_\ell) &= W_{\ell}^{jk}W_{\ell}^{mn}\operatorname{Cov}(\hat{C}^{jk}_\ell,\hat{C}^{mn}_\ell) \nonumber \\
				    &= W_{\ell}^{jk}W_{\ell}^{mn} \frac{2}{N_p} C_\ell^{jm} C_\ell^{kn} \; ,
					 \label{eq:gaussian_variance_estimate}
\end{align}
where we additionally assumed Gaussianity of the map and applied Wick's theorem. To be accurate, the derivation of the standard SpILC weights only requires the assumption of \emph{noise} Gaussianity (see App.\ \ref{sec:gaussianity} for a justification).

To derive $W_{\ell}^{ij}$ which minimizes $\operatorname{Var}(\hat{K}_\ell)$ subject to  $W_{\ell}^{ij}a^ia^j = 1$, we use Lagrange's method of undetermined multipliers to yield the set of $M+1$ simultaneous equations, where $M\equiv N(N+1)/2$, and $N$ is the number of frequency channels: 
 \begin{align}
	\begin{cases}
		\partial_{W^{ij}_\ell} \left[W_{\ell}^{ab}W_{\ell}^{cd}C_\ell^{ac}C_\ell^{bd} - \lambda(W_{\ell}^{cd}a^ca^d-1) \right] =0 & \mbox{, $i\leq j$}\\
		W_{\ell}^{cd} a^c a^d = 1 \; ,
	\end{cases} \; 
	\label{eq:pilc_eqs}
\end{align}
as the symmetric property of $W_{\ell}^{ij}$ implies that only the weights $W_{\ell}^{ij}$ with  $i\leq j$ are independent degrees of freedom. Labeling these $M$ degrees of freedom with Greek indices  $\mu=\{1,\dots,M\}$, each corresponding to a pair  $(i,j)$,  $i\leq j$, The weights are solved to give
\begin{align}
	w^\mu_\ell = \frac{D^{-1}_{\ell,\mu\nu}t_\nu}{t_{\rho}D^{-1}_{\ell,\rho\tau}t_\tau} \; ,
	\label{eq:pilc_weights}
\end{align}
where $t_\mu$ is a vector
\begin{align}
	\mathbf{t}\equiv 
	\underbrace{\begin{pmatrix}
			a^1a^1 &a^1 a^2  & \cdots & a^1 a^N & a^2a^2 & a^2 a^3 & \cdots & a^Na^N
	\end{pmatrix}^{\text{T}}}_{\text{$N(N+1)/2$ rows}} \; ,
	\label{eq:pilc_def_t}
\end{align}
and the matrix $D_\ell^{\mu\nu}$ can be found in Eq.\ (\ref{eq:pilc_def_D}). The weight vector is defined as follows:
\begin{align}
	\mathbf{w}_\ell &\equiv 
	\underbrace{\begin{pmatrix}
			\bar{W}_{\ell}^{11} &\bar{W}_{\ell}^{12} & \cdots & \bar{W}_{\ell}^{1N}&\bar{W}_{\ell}^{22} & \bar{W}_{\ell}^{23} \cdots \bar{W}_{\ell}^{NN}
	\end{pmatrix}^{\text{T}}}_{\text{$N(N+1)/2$ rows}} \; ,
\label{eq:pilc_def_w}
\end{align}
where
\begin{align}
	\bar{W}_{\ell}^{ij} \equiv (2-\delta_{ij})W_{\ell}^{ij}=
	\begin{cases}
		W_{\ell}^{ij}\; , &\mbox{for $i=j$}\\
		2W_{\ell}^{ij}\; , &\mbox{for $i\neq j$}
	\end{cases}\; .
	\label{eq:pilc_def_barW}
\end{align}
See App.\ \ref{app:derivation_spilc} for a detailed derivation. 

In practical applications, we have access only to a single realization of the power spectra $\hat{C}^{ij}_\ell$---that of the observed sky, so we derive the weights and estimators only from one sky realization through \emph{internal} combination.
\begin{align}
	\hat{K}_\ell^\text{1real} \equiv \hat{W}_{\ell,ij}^\text{1real}\hat{C}^{ij}_\ell \; .
	\label{eq:oneskyrealization}
\end{align}
Although our derivations are concerned with $\hat{K}_\ell$, we will mainly discuss simulation results for the estimator  $\hat{K}^\text{1real}_\ell$.
In this case we minimize the sample variance in that realization:
\begin{align}
	\operatorname{Var}(\hat{K}_\ell^{\text{1real}}) &\approx \hat{\operatorname{Var}}(\hat{K}_\ell^{\text{1real}})=\hat{W}^{\text{1real}}_{\ell,jk}\hat{W}^{\text{1real}}_{\ell,mn} \frac{2}{N_p} \hat{C}^{jm}_\ell \hat{C}^{kn}_\ell \; ,
	\label{eq:oneskyrealization_variance}
\end{align}
where the weights keep the same functional form, but with $C_\ell^{ij}$ replaced with $\hat{C}_\ell^{ij}$.

Having finite samples mean that the latter may suffer from expectation bias (analogous to the map-level ILC bias) due to chance correlations between signal and noise, making $\langle \hat{K}_\ell^\text{1real}  \rangle\neq\langle \hat{K}_\ell  \rangle$. Writing out the bias,
 \begin{align}
	 \langle \hat{\varepsilon}_\ell  \rangle &= \langle \hat{K}_\ell-C^{ss}_\ell  \rangle = \langle \hat{K}_\ell  \rangle - C^{ss}_\ell \nonumber \\
					    &= \langle W_{\ell}^{ij} \hat{C}^{ij}_\ell  \rangle - C_\ell^{ss} \nonumber \\
					    &= \langle W_{\ell}^{ij} (a^i s_p + n^i_p)(a^j s_p^* + n^{j,*}_p)  \rangle - C_\ell^{ss} \nonumber \\
					     &= \langle W_{\ell}^{ij} n^i_p n^{j,*}_p  \rangle + 2 \langle W_{\ell}^{ij} a^i s_p n^{j,*}_p  \rangle  \; .
					    \label{eq:noise_bias}
\end{align}
The bias $\langle W_{\ell,ij} n^i_p n^j_p  \rangle$ is the usual \emph{noise bias} in power spectrum methods, as any non-zero diagonal entries of $W_{\ell}^{ij}$ would contribute noise auto-correlation into the resulting spectra, even if there is no noise correlations between channels. The bias $\langle W^{ij}_\ell a^i s_p n^{j,*}_p  \rangle$ vanishes in the ensemble average, however the sample covariance due to chance correlation between the noise and the signal $N^{-1}_p\sum_p s_p n^{j,*}_p$ is non-vanishing and contributes to the expectation bias $\langle \hat{W}^{\text{1real}}_{\ell,ij} a^i s_p n^{j,*}_p  \rangle$. 
 The magnitude of the expectation bias will be discussed through simulations in Sec.\ \ref{sec:5}.

\subsection{Equivalence between standard ILC and SpILC}
\label{sec:2c}
The standard ILC and SpILC weights for auto- and cross-spectra are identical, i.e.\  the weights $w_{\ell,i}^\text{ILC}$ for the ILC maps  $\hat{s}_p\equiv w_{\ell,i}^{\text{ILC}}x^i_p$ satisfy
\begin{align}
	W^\text{SpILC}_{\ell,ij} = w^\text{ILC}_{\ell,i} w^\text{ILC}_{\ell,j} \; .
\end{align}
 This arises as the one-component standard ILC and SpILC constrained-optimization problems turn out to be equivalent---App.\ \ref{app:eqv_ilc_pilc} provides a proof for this argument.
This means that the spectra estimated from ILC maps $\hat{s}_p$ equals our standard SpILC estimator: 
\begin{align}
	\hat{K}_\ell^{\text{SpILC}} &= W_{\ell,ij}^{\text{SpILC}}\hat{C}_\ell^{ij} = w_{\ell,i}^{\text{ILC}}w_{\ell,j}^{\text{ILC}} \frac{1}{N_p}\sum_{p=1}^{N_p} x^i_p x^{*,j}_p \nonumber \\
			       &= \frac{1}{N_p}\sum_{p=1}^{N_p} \hat{s}_p^\text{ILC} \hat{s}_p^{*,\text{ILC}} \; .
\end{align}
However, we will find in the next section that the equality no longer holds between constrained ILC and constrained Spectral ILC (cSpILC) when we extend to a multiple-component model and impose deprojection constraints (Sec.\ \ref{sec:cSpILC}), \emph{if} some constraints can be dropped as a result of zero spatial correlation between some components.

\section{Constrained S\lowercase{p}ILC (\lowercase{c}S\lowercase{p}ILC)}
\label{sec:cSpILC}
This section holds the key analytical result of our study, where we show that the constrained SpILC (cSpILC) method can produce spectrum estimators with lower variance than cILC. We demonstrate this with the two-component model
 \begin{align}
 	x^i_p &= a^i s_p + b^i y_p + n^i_p \; ,
 \end{align}
 which can be readily generalized to arbitrarily many components.
 We impose that the estimator has unit response to the spectrum of interest ($ss$-spectrum here), and zero response to all other spectra:
\begin{align}
	W_{\ell}^{ij}\langle  (a^i s_p + b^i y_p)(a^j s^*_p + b^j y^*_p) \rangle &= \langle  s_p s^*_p  \rangle \; .
\end{align}
Generalizing from the constrained ILC (cILC) method, we can impose the following \emph{normalization} and \emph{deprojection} constraints, which we call collectively as the \texttt{strong}-cSpILC constraints:
 \begin{align}
	 \begin{cases}
	 W_{\ell}^{ij}a^i a^j = 1 \; ,\\
	 W_{\ell}^{ij} b^i b^j = 0 \; , \\
	 W_{\ell}^{ij} a^{(i}b^{j)} = 0 \; ,
 \end{cases} 
 \; \; \text{(\texttt{strong}-cSpILC constraints)}
\end{align}
where the symmetrization bracket is defined with $a^{(i}b^{j)}=(a^ib^j + a^j b^i)/2$.
However, if we know \emph{a priori} that the two components $s_p$ and  $y_p$ have zero cross spectra, $\langle  s_p y^*_p  \rangle=0$, then 
we need only to impose weaker constraints for our purpose:
 \begin{align}
	 \begin{cases}
	 W_{\ell}^{ij}a^i a^j = 1 \; ,\\
	 W_{\ell}^{ij} b^i b^j = 0 \; , \\
	 \end{cases}
 \; \; \text{(\texttt{weak}-cSpILC constraints)}
 \label{eq:weak_cspilc}
\end{align}
which we call the \texttt{weak}-cSpILC constraints.

In App.\ \ref{app:eqv_cilc_cpilc} we generalize the ILC-SpILC equivalence of the last subsection and show that \texttt{strong}-cSpILC method produces the same weights as cILC, and \textbf{as a result \texttt{weak}-cSpILC must obtain an equal or smaller variance than cILC due to its fewer constraints.} In Sec.\ \ref{sec:3} we show through simulations that \texttt{weak}-cSpILC indeed achieves lower variance than cILC.
This result demonstrates that the variance reduction of \texttt{weak}-cSpILC compared to cILC does not come directly from the increased degrees of freedom from $N$ to $N(N+1)/2$ of the weights, as the number of constraints increases correspondingly in \texttt{strong}-cSpILC and the same variance as cILC is obtained. Rather, it is the \emph{explicit relaxation of constraints} in Eq.\ (\ref{eq:weak_cspilc}) enabled by the SpILC parameterization at the spectral level that contributes to its variance reduction.

The weights for \texttt{weak}-cSpILC are derived by solving
\begin{align}
	 &\begin{cases}
		 \partial_{W_\ell^{ij}} \left[W_{\ell}^{ab}W_{\ell}^{cd}C^{ac}_\ell C_\ell^{bd} - \lambda(W_{\ell}^{cd}a^ca^d-1) \right. \\
		 \phantom{asdasdasdasdaasasddaaaaa}\left.-\mu W_{\ell}^{cd}b^c b^d\right]  =0 \;\;, i\leq j \\
	 W_{\ell}^{cd}a^c a^d = 1 \\
	 W_{\ell}^{cd} b^c b^d = 0 
 \end{cases} \; ,
\end{align}
where $\lambda$ and $\mu$ are Lagrange multipliers, giving the \texttt{weak}-cSpILC weights
 \begin{align}
	 w^\mu_\ell &= \frac{\left( \mathbf{u}^\text{T}\mathbf{D}^{-1}_\ell\mathbf{u} \right) D^{-1}_{\ell,\mu\rho} t^\rho-\left(\mathbf{t}^\text{T}\mathbf{D}_\ell^{-1}\mathbf{u}\right) D^{-1}_{\ell,\mu\rho}u^\rho}{\left(\mathbf{t}^\text{T}\mathbf{D}_\ell^{-1}\mathbf{t}\right)\left( \mathbf{u}^\text{T}\mathbf{D}_\ell^{-1}\mathbf{u} \right)- \left(\mathbf{t}^\text{T}\mathbf{D}_\ell^{-1}\mathbf{u}\right)^2} \; ,
	 \label{eq:cspilc_weights}
\end{align}
where the Greek indices ranges over the $N(N+1)/2$ degrees of freedom of pairs of  $(i,j)$,  $i\leq j$, as in Eq.\ (\ref{eq:pilc_weights}). The vector $u^\mu$ is defined as
\begin{align}
	\mathbf{u}\equiv 
	\underbrace{\begin{pmatrix}
			b^1b^1  & b^1 b^2\cdots &b^1b^N & b^2b^2 & b^2 b^3 & \cdots & b^Nb^N
	\end{pmatrix}^{\text{T}}}_{\text{$N(N+1)/2$ rows}} \; ,
\end{align}
and the vector $t^\mu$ (Eq.\ (\ref{eq:pilc_def_t})) is similarly defined with $b\mapsto a$. The matrix $D_\ell^{\mu\nu}$ can be found in Eq.\ (\ref{eq:pilc_def_D}). Note the definition of the weight vector $w_\ell^\mu$ defined in Eqs.\ (\ref{eq:pilc_def_w}) and (\ref{eq:pilc_def_barW}).
See App.\ \ref{app:derivation_cspilc} for a detailed derivation.

Unlike the standard SpILC where only the Gaussianity of noise is assumed and \emph{not} the components, for \texttt{weak}-cSpILC there are non-vanishing connected four-point functions
\begin{align}
    \langle s_p y_p^* s_q y_q^* \rangle_c &= 0 \; ,\\
    \langle s_p s^*_p s_q y_q^*\rangle_c &= 0 \; ,
\end{align}
where the components $s_p$ and $y_p$ are those assumed to be uncorrelated in their two-point functions. Therefore, the weight derivation for \texttt{weak}-cSpILC using Wick's theorem implicitly imposes an additional assumption that the specific four-point functions above vanishes. If this additional assumption is violated, the \texttt{weak}-cSpILC estimator will be sub-optimal. See Appendix \ref{sec:gaussianity} for details.

Generalization to multiple components is straightforward. Suppose we have a component $k_p$ with SED $c^i$ which we also want to deproject, and we know \emph{a priori} that  $s_p$ is spatially uncorrelated with both $y_p$ and $k_p$, while allowing spatial correlation between $k_p$ and $y_p$. The additional constraints we need to include would be
\begin{align}
	W_{\ell}^{ij} c^i c^j &= 0 \; , \\
	2W_{\ell}^{ij} c^{(i}b^{j)} &= 0 \; .
\end{align}
This is the physical example we consider for simulations in the following section: $s_p$ is the CMB+kSZ (both components have the blackbody SED), $y_p$ and  $k_p$ are the tSZ and the cosmic infrared background (CIB) map respectively, where tSZ and CIB are spatially correlated, and both are uncorrelated with the CMB+kSZ. Writing the SED for component $s$ as $f^i_s$, the constraints for estimating the (CMB+kSZ)$\times$(CMB+kSZ) spectrum are
\begin{align}
	\begin{cases}
		 W_{\ell}^{jk}f^j_\text{CMB+kSZ} f^k_\text{CMB+kSZ} = 1 & \mbox{((CMB+kSZ)$^2$)} \\
		 W_{\ell}^{jk}f^j_\text{tSZ} f^k_\text{tSZ} = 0 & \mbox{(tSZ$\times$tSZ)} \\
		 W_{\ell}^{jk}f^j_\text{CIB} f^k_\text{CIB} = 0 & \mbox{(CIB$\times$CIB)} \\
		 2W_{\ell}^{jk} f^{(j}_\text{tSZ} f^{k)}_\text{CIB} = 0 & \mbox{(tSZ$\times$CIB)} 
	\end{cases} \; .
\end{align}

To estimate another spectrum, e.g.\ tSZ$\times$tSZ, the RHS of the constraint of that spectrum is simply replaced with 1 (e.g.\ $ W_{\ell}^{jk}f^j_\text{tSZ} f^k_\text{tSZ}=1$) and the rest with 0;  explicitly, tSZ$\times$tSZ is normalized while (CMB+kSZ)$\times$(CMB+kSZ), CIB$\times$CIB and tSZ$\times$CIB are deprojected.
\section{Data-split S\lowercase{p}ILC and \lowercase{c}S\lowercase{p}ILC}
\label{sec:3}

As we saw in Eq.\ (\ref{eq:noise_bias}), the noise term contributes variance to the estimator $\hat{K}_\ell$, leading to a noise bias.
The noise bias due to instrumental and atmospheric noise---which is assumed to be uncorrelated across channels at different times---can be removed using \emph{data splits} (defined below). The residual noise bias would be contributed by unmodeled components, which are correlated across channels.

Suppose $x^i_p$ is a map constructed from data collected from time $t_0$ to  $t_0 + \Delta t$. Two noisier maps $x^{i,\ds}_p$ and $x^{i,\dss}_p$, referred to as \emph{data splits}, can be constructed from the time segments $[t_0,t_0+\Delta t/2]$ and  $[t_0+\Delta t/2,t_0+\Delta t]$ respectively.  The two time segments are labeled 1 and 2 respectively. 

We can build the following SpILC estimator with vanishing \emph{instrumental} noise bias, as same-split, same-channel spectra are removed:
\begin{align}
	\hat{K}_\ell^\text{DS} &= \sum_{ij}^N W_{\ell}^{ij} \left[\delta_{ij}\hat{C}^{ij}_{\ell,(\ds\dss)} + (1-\delta_{ij})\hat{C}_\ell^{ij}\right] \\
	 &= \sum_{ij}^N W_{\ell}^{ij} \left[\delta_{ij}\hat{C}^{ij}_{\ell,\ds\dss} + (1-\delta_{ij})\hat{C}^{ij}_\ell\right] \; ,
     \label{eq:data_split1}
\end{align}
where
\begin{align}
	\hat{C}^{ij}_{\ell,12} &\equiv \frac{1}{N_p}\sum_p x^i_{p,1}x^{j,*}_{p,2} \; \; , \; \; 
	\hat{C}^{ij}_{\ell,21} \equiv \frac{1}{N_p}\sum_p x^i_{p,2}x^{j,*}_{p,1} \; .
\end{align}
This estimator will have lower variance than the estimator $\hat{K}^{\text{DS2}}_{\ell,\ds\ds} = W_{\ell}^{ij}\hat{C}^{ij}_{\ell,\ds\dss}$, as the construction of the latter discarded the information in the same-split, different-channel spectra $(1-\delta_{ij})\hat{C}^{ij}_{\ell,\ds\ds}$ and  $(1-\delta_{ij})\hat{C}^{ij}_{\ell,\dss\dss}$. 
We note the estimator $\hat{K}^\text{DS0}_\ell$ instead of $\hat{K}^\text{DS}_\ell$ is required to remove bias from correlated atmospheric noise across channels, which is currently ignored in the variance comparison (in simulations we set the ensemble noise cross-spectra to be zero, $N^{ij}=0$ for $i\neq j$). The weights $W_{\ell}^{ij}$ are determined as usual by minimizing subject to constraints on the variance of $\hat{K}_\ell$, which is derived in App.\ \ref{app:splits_c} for both $\hat{K}^{\text{DS0}}_\ell$ and $\hat{K}^{\text{DS}}_\ell$.

\section{Simulations}
\begin{figure}
	\includegraphics[width=0.98\linewidth]{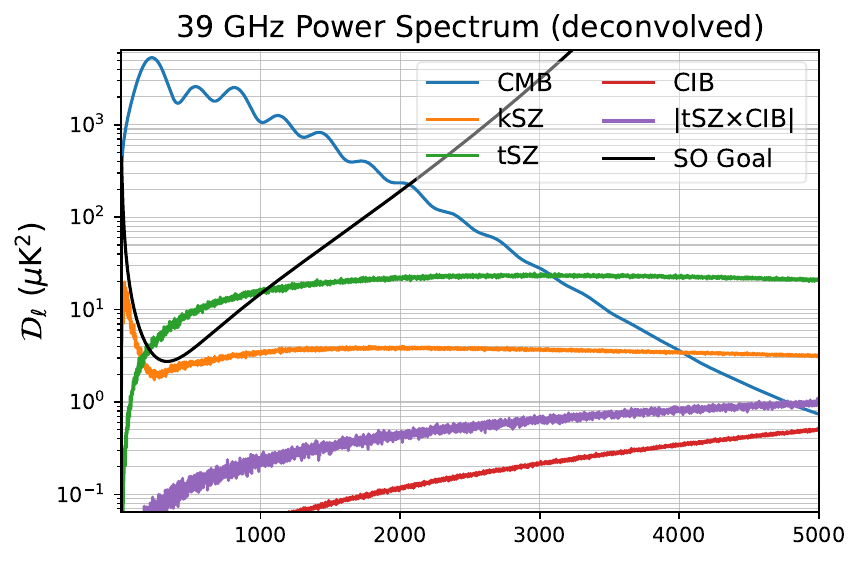}
	\includegraphics[width=0.98\linewidth]{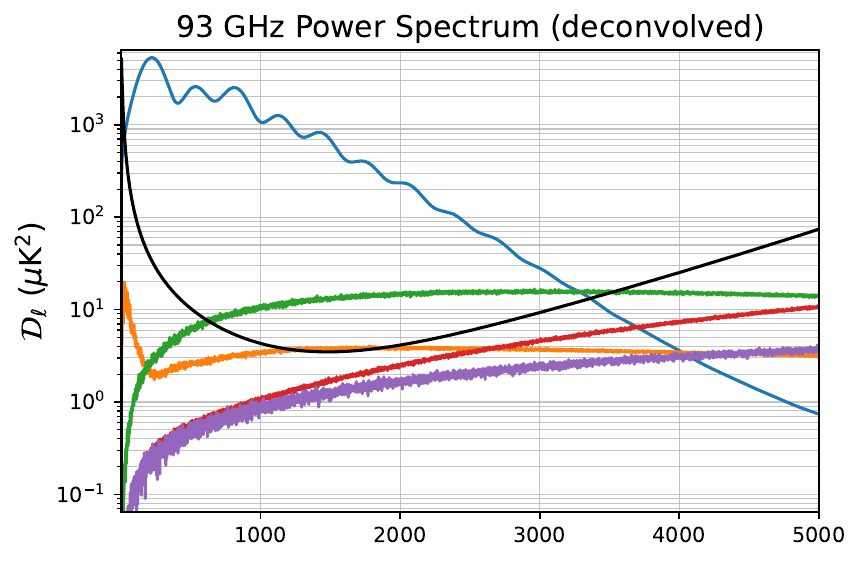}
	\includegraphics[width=0.98\linewidth]{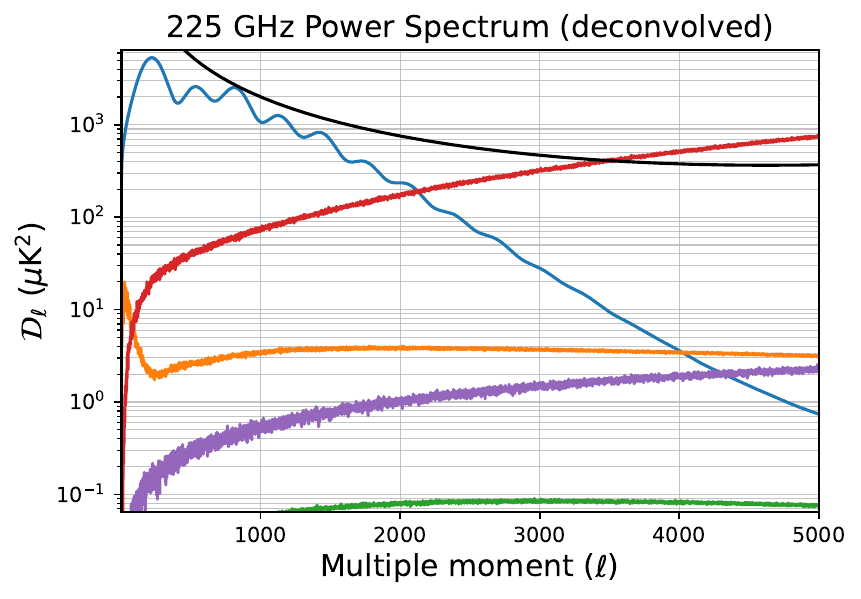}
	\caption{Auto and cross spectra $\mathcal{D}_\ell \equiv\ell(\ell+1)C_\ell/2\pi$ for different components of the simulated maps at 39 GHz (top), 93 GHz (middle), and 225 GHz (bottom). The simulated data are beam-deconvolved with beams and noise profiles corresponding to SO LAT goal level.
}
	\label{fig:spectra}
\end{figure}
Henceforth, when we use the term ``SpILC'' we are referring only to \texttt{weak}-cSpILC, noting that any discussions on \texttt{strong}-cSpILC (one-component SpILC) can be replaced by cILC (ILC).

\begin{figure*}
	\includegraphics[width=0.48\linewidth]{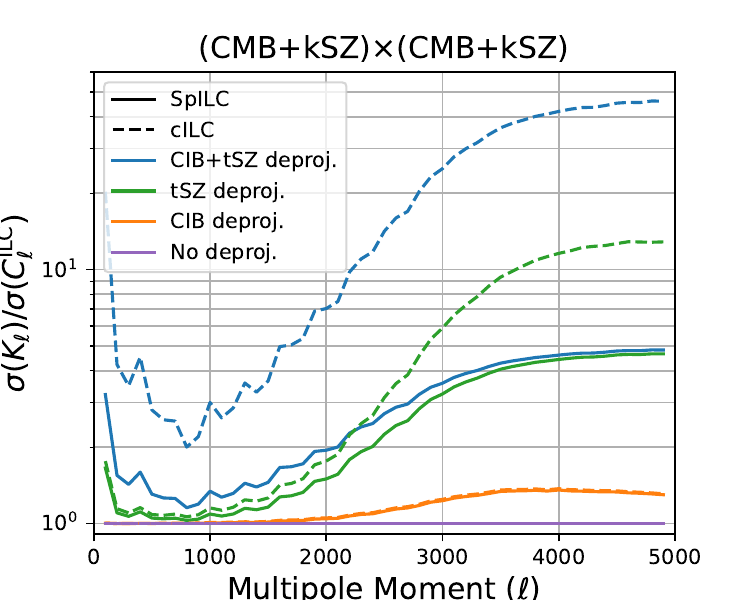}
	\includegraphics[width=0.48\linewidth]{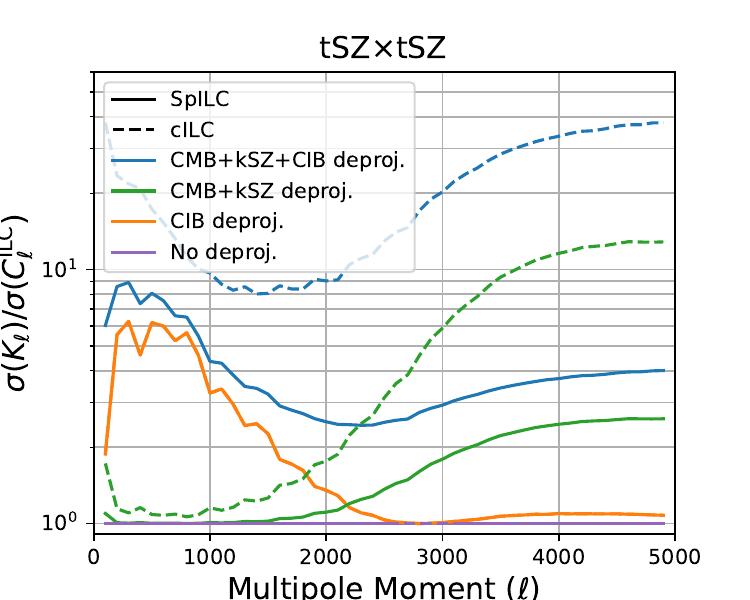}
\caption{\textit{Left}: Errorbar size ratios of various SpILC (solid) and cILC estimators (dashed) of the CMB+kSZ power spectrum compared to the ILC (no deprojection) power spectrum. The line colors denote the corresponding deprojections: tSZ$\times$tSZ + tSZ$\times$CIB + CIB$\times$CIB (blue), CIB$\times$CIB + tSZ$\times$CIB (orange), tSZ$\times$tSZ + tSZ$\times$CIB (green), and no deprojections (purple) where cILC reduces to standard ILC. At small scales $\ell\gtrsim 4000$, the CIB and tSZ deprojected SpILC (cILC) errorbar $\sigma(\hat{K}_\ell)$ is more than 5 (40) times larger than the ILC errorbar size $\sigma(\hat{C}_\ell^\text{ILC})$, and in turn the cILC estimator has $\gtrsim 8$ times the errorbar size of the SpILC estimator. 
\textit{Right}: Similar to left, but for the tSZ power spectrum, where the fully constrained case deprojects CMB+kSZ and CIB.}
	\label{fig:var_comparison}
\end{figure*}

\begin{figure*}
	\includegraphics[width=\linewidth]{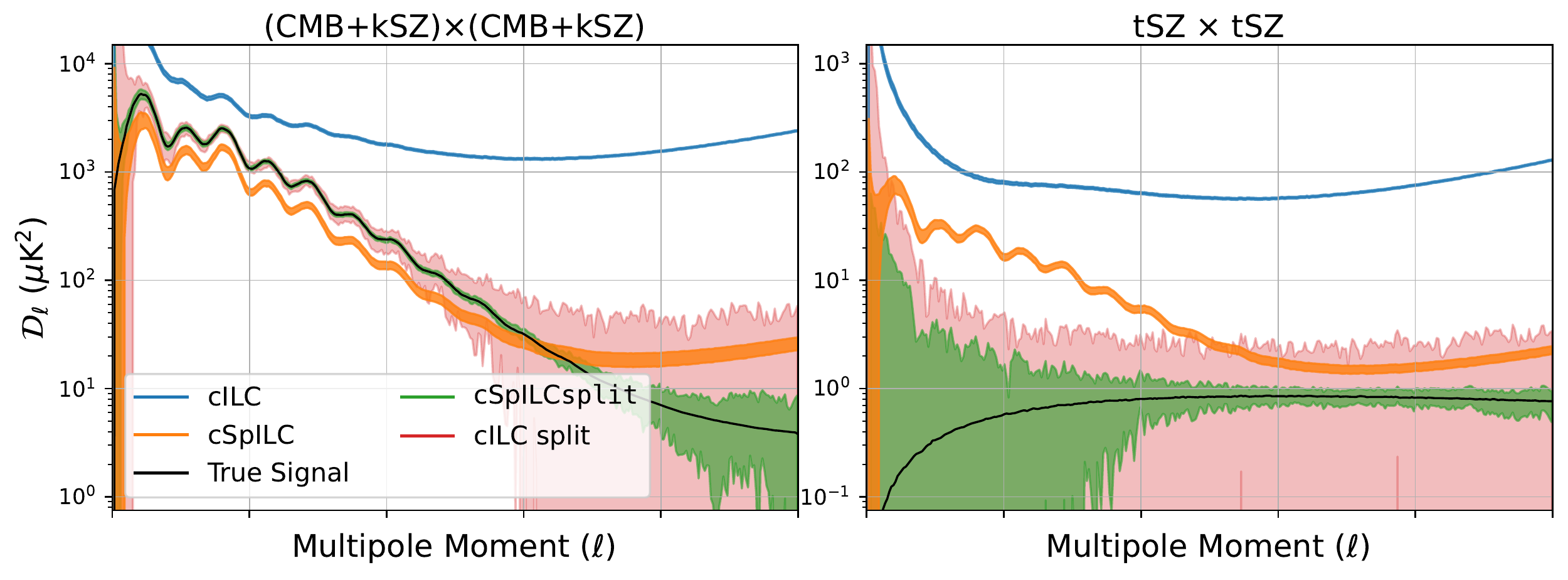}
	\caption{The true (black)  CMB power spectrum $\mathcal{D}_\ell^{\text{CMB+kSZ}}\equiv \ell(\ell+1)C^{\text{CMB+kSZ}}_\ell/2\pi$ (left panel) and tSZ power spectrum $\mathcal{D}_\ell^{yy}$ (right panel) are shown along with the $\pm1$$\sigma$ regions of the estimator  $\hat{K}_{\ell}^\text{1real}$ for the constrained ILC (cILC, blue), constrained Spectral ILC (SpILC, orange), the data-split constrained ILC (cILC\texttt{split}, red) and the data-split constrained Spectral ILC (SpILC\texttt{split}, green). The deprojected components are the tSZ and CIB for the CMB spectrum (left), and CMB+kSZ and CIB for the tSZ spectrum (right). At each multipole moment, the measured spectra $\hat{C}^{ij}_\ell$ are band-averaged with a band-width of $\Delta\ell=30$ as in Eq.\ (\ref{eq:bandavg}). The unbiased estimator with the smallest variance is the SpILC\texttt{split}, and it has significantly lower variance than the analagous unbiased map-version, the cILC\texttt{split} (as quantified in Figs.\ \ref{fig:var_comparison} and \ref{fig:one_real_6noise}). The large biases for cILC and SpILC are due to noise bias, as the split estimators are unbiased.} 
	\label{fig:temp_3comp_12noise_19real_ensemble}
\end{figure*}
\subsection{Description of Simulations}
\label{sec:4}

We simulate beam-deconvolved maps in \emph{linearized differential thermodynamic units}, where the SED of CMB and kSZ temperature anisotropies $\Delta T_p^\text{CMB}$ and $\Delta T_p^\text{kSZ}$ is unity for all channels $i$ (no summation implied):
 \begin{align}
	  x^i_p \equiv \Delta T_p^\text{CMB} +\Delta T_p^\text{kSZ} + f_\text{tSZ}^i y_p + f^i_\text{CIB} I_{p,\text{CIB}}^{143} + B_{i}^{-1}\otimes n^i_p \; ,
\end{align}
where the Compton-$y$ parameter is  $y_p$, the intensity of CIB at 143 GHz is  $I^{143}_{p,\text{CMB}}$,
$B^i$ is the Gaussian beam for channel $i$ with full width at half maximum (FWHM) specified for SO LAT (Table 1 of Ref.\ \cite{so}), $\otimes$ is the convolution operation, and the noise $n^i_p$ is Gaussian and uncorrelated across channels (and splits) following SO goal noise spectra generated using public package \texttt{so\_noise\_models
} \texttt{v3.1.2} \cite{so,so_noise_model}), which includes both large-scale atmospheric noise and white instrumental noise. The large-scale atmospheric noise correlations across channels are currently ignored in this study.

The maps  $\Delta T_p^\text{CMB}, \Delta T_p^\text{kSZ}, y_p, I^\text{143}_{p,\text{CIB}}$ are simulated to be Gaussian by drawing random realizations from respectively the TT power spectrum from best-fit $\Lambda$CDM parameters to ACT DR4 data \cite{act_dr4_maps,choi2020}, Websky kSZ (both $z<4.5$ and $z>5.5$ patchy reionization) power spectrum, and tSZ, CIB power and cross spectra estimated from the WebSky tSZ and CIB (143 GHz) maps \cite{websky}. The simulated tSZ and CIB maps are correlated. Spatial decorrelation of the CIB is ignored by scaling the CIB map at 143 GHz to simulate CIB maps from other frequency channels. 
We use the full sky geometry, and tested that changing the map geometry to a partial sky region (such as the ACT Deep56 region of 834 deg$^2$ of sky) does  not change our results, particularly the fractional improvement of errorbar sizes.

\begin{figure*}
	\includegraphics[width=0.49\linewidth]{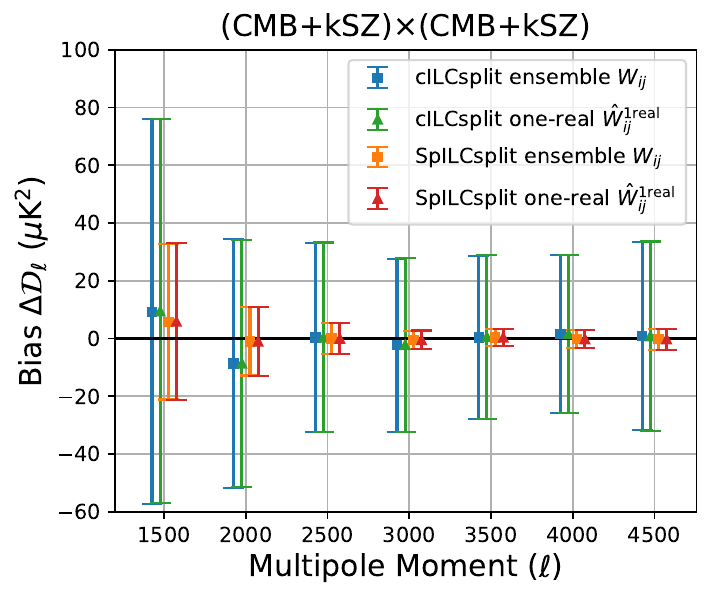}
	\includegraphics[width=0.49\linewidth]{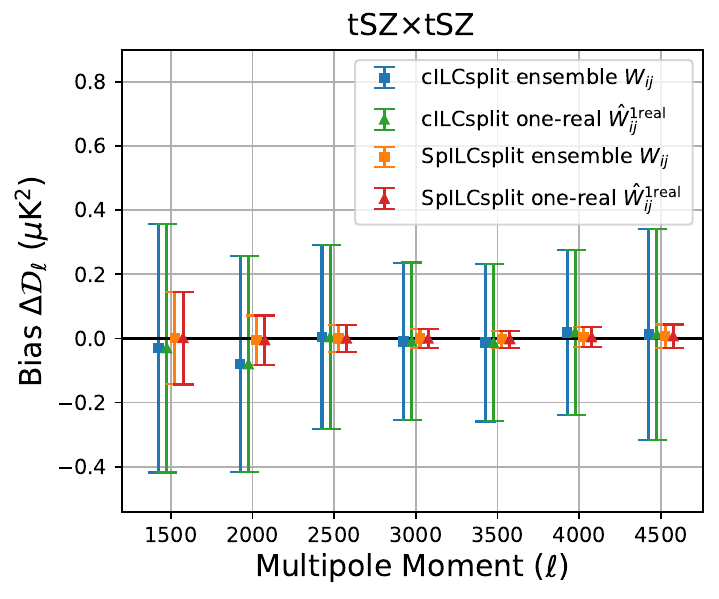}\\
	\caption{  \textit{Left:} Biases of (CMB+kSZ)$\times$(CMB+kSZ) power spectrum estimated at SO goal noise levels for cILC\texttt{split} and SpILC\texttt{split},  which eliminates instrumental noise bias.  The errorbars are $\pm1$ $\sigma$ about the mean error  $\langle \hat{\varepsilon}_\ell  \rangle$ over 100 realizations, where the error is defined as the difference of the estimator value from the true spectra. For each estimator, the weights are estimated using two methods: 1.\ weights $W_{\ell}^{ij}$ (blue for cILC\texttt{split}, orange for SpILC\texttt{split} from the realization-averaged spectra $\langle \hat{C}_\ell^{ij}  \rangle$, whereas 2.\ weights $\hat{W}^{\text{1real}}_{\ell,ij}$ are derived for each realization using data from that realization (green for cILC\texttt{split}, red for SpILC\texttt{split}). The tSZ and CIB are deprojected, and the measured spectra for each realization are band-averaged with a band-width of $\Delta\ell=30$.  \textit{Right:} Similar to the left panel, but for the tSZ power spectrum. The CMB+kSZ and CIB are deprojected.
}
	\label{fig:one_real_6noise}

\end{figure*}

We assume exactly known SEDs. Here $f^i_\text{tSZ}$ is the SED of tSZ in thermodynamic units (dimension of temperature), which is obtained analytically as \cite{sunyaev72, kompaneets}:
 \begin{align}
 f^\nu_\text{tSZ} &= \bar{T}\left(X\coth\left( \frac{X}{2} \right)-4\right) \; ,
\end{align}
where $X\equiv h\nu/k\bar{T}$, $\bar{T}$ is the temperature of the CMB monopole, $\nu$ indicates frequency, and $h$ and $k$ are Planck's constant and Boltzmann's constant respectively.
The CIB is modeled as a modified blackbody, the SED (dimension of temperature/intensity) of the CIB intensity map is
\begin{align}
	f_\text{CIB}^\nu \propto \frac{\nu^{3+\beta}}{e^{X_\text{CIB}}-1}\times\frac{e^X-1}{Xe^X}\frac{\bar{T}}{M^\nu(\bar{T})} \; ,
	\label{eq:fcib}
\end{align}
where $X_\text{CIB}\equiv h\nu/kT_\text{CIB}$, with $T_\text{CIB}$ the effective temperature of the CIB and $\beta$ its spectral index. We use $T_\text{CIB}=10.70$ K, $\beta=1.7$ \cite{ymap_coulton}. The SED is normalized at 143 GHz.

In summary, each data realization consists of a linear combination of simulated CMB, kSZ, tSZ and CIB maps, weighted by their SEDs, and SO-goal level beam-deconvolved noise.
The noise level in each split is larger than un-split maps by a factor of $\sqrt{2}$ at map level. We show the power spectra of the signals at several frequencies in Fig.\ \ref{fig:spectra}.  Finally, we perform band-averaging to the measured spectra $\hat{C}^{ij}_\ell$ by
\begin{align}
    \hat{C}^{ij}_{\ell_0,\text{band-avg}} &= N_\text{mode}^{-1}\sum_{\ell=\ell_0-\Delta\ell/2}^{\ell_0+\Delta\ell/2} (2\ell+1)\hat{C}^{ij}_\ell \; ,
    \label{eq:bandavg}\\
  N_\text{mode} &\equiv  \sum_{\ell=\ell_0-\Delta\ell/2}^{\ell_0+\Delta\ell/2}(2\ell+1) \; ,
\end{align}
with a band-width of $\Delta\ell=30$ centered at $\ell_0$. 

\subsection{Simulation Results}
\label{sec:5}
For cILC, cILC\texttt{split}, SpILC, and SpILC\texttt{split}, we apply for each realization our estimator $\hat{K}^\text{1real}_\ell$ (see Eq.\ \eqref{eq:oneskyrealization} and \eqref{eq:oneskyrealization_variance}) to estimate the (CMB+kSZ)$\times$(CMB+kSZ) and tSZ$\times$tSZ spectra. The spectra and bias are plotted in Fig.\ \ref{fig:temp_3comp_12noise_19real_ensemble} and compared to the truth (ACT best-fit CMB$\times$CMB + measured Websky kSZ$\times$kSZ spectra, and measured Websky tSZ$\times$tSZ spectra), whereas the variances are compared in Fig.\ \ref{fig:var_comparison}.

\textbf{Variance.} We first present in Fig.\ \ref{fig:var_comparison} the comparison of estimator variances. We plot the errorbar ratio of various cILC (dashed lines) and SpILC\texttt{split} (solid lines) estimators with different deprojection choices (denoted by colors) to the standard ILC estimator. As data split estimators discard same-channel, same-split auto spectra, the variance is slightly larger than the non-split SpILC (15\% at small scales).

Impressively, for both (CMB+kSZ)$\times$(CMB+kSZ) and tSZ$\times$tSZ spectra at small scales, the errorbar size of the SpILC estimator can be significantly smaller than that of the cILC estimator. 
At $\ell\gtrsim 4000$, the fully-deprojected SpILC estimator has around 7 times smaller errorbar sizes than the fully-deprojected cILC, and even a 2 times smaller errorbar than the just-tSZ (just-CMB+kSZ) deprojected cILC, despite SpILC deprojecting also the CIB component.
As expected, the smallest errorbars are achieved by the no-deprojection cILC, which is equivalent to the standard harmonic ILC. For $\ell\gtrsim 4000$, it is more than 30 times smaller than the fully-deprojected cILC.

The large improvement in errorbar sizes at small scales where instrumental noise is significant suggests two particular use cases where the application of SpILC would be particularly interesting: 1.\ when noise power is significant compared to signal power, e.g.\ $B$-mode detection, and 2.\ when there are many components to model, e.g.\ using a moment expansion of the CIB where the CIB SED is represented as a Taylor expansion about parameters $\beta, T_\text{CIB}$ (Eq.\ (\ref{eq:fcib})) \cite{chluba_moment} and deprojecting the zeroth and first order moments (and optionally higher moments) \cite{cmilc}, where the resulting cILC spectra variances become high.

\textbf{Noise Bias.} From Fig.\ \ref{fig:temp_3comp_12noise_19real_ensemble}, one sees that the cILC and SpILC estimator are biased away from the true spectra (black lines). The bias in SpILC is smaller than that in cILC, and exhibits different angular behaviour at small scales: 1.\ unlike cILC, the instrumental noise bias $\sum_i W_{{\ell}}^{ii}\hat{N}^{ii}_\ell$ is not necessarily positive since weights for auto spectra can be negative; 2.\ one sees acoustic features in the CMB-deprojected tSZ$\times$tSZ SpILC spectrum since terms proportional to $\hat{C}^{ss}_\ell$ remain in the variance in the SpILC constrained-optimization problem, such as $W_{\ell}^{jk}W_{\ell}^{mn}a^j a^m\hat{C}_\ell^{ss}\hat{N}_\ell^{kn}$ and $W_{\ell}^{jk}W_{\ell}^{mn}a^j a^m b^k b^n \hat{C}_\ell^{ss}\hat{C}^{yy}_\ell$. 

By implementing data splits for SpILC which eliminate instrumental noise bias, one sees that the SpILC\texttt{split} and cILC\texttt{split} estimators are unbiased. As the data splits are expected to remove all noise bias in our simulation (as we have only included uncorrelated noise between frequency channels---even for the atmospheric part, and all sky components are assumed to be modeled), and no bias remains in the SpILC\texttt{split} estimator, we conclude that the aforementioned bias and behaviours are solely attributed to noise bias. Atmospheric noise that correlates between different channels can be further mitigated by constructing a data-split estimator only with cross-spectra $\hat{C}_{\ell,12}^{ij}, \hat{C}_{\ell,21}^{ij}$, even for $i\neq j$.

 \textbf{Bias due to single sky realization.} The weights $\hat{W}_{ij}$ measured from one sky realization of data $\hat{C}_\ell^{ij}$ will not match weights $W_{\ell}^{ij}$ calculated from ensemble-averaged $C_\ell^{ij}$, as empirical covariances $\hat{C}_\ell^{ij}$ measured from finite sample size deviate from the true covariances, e.g.\ from chance correlations between components and noise. As a result, $\hat{K}_\ell^\text{1real}$ may suffer from \emph{expectation biases} due to statistical fluctuations of $\hat{C}_\ell^{ij}-C_\ell^{ij}$ from zero. Note that map-level ILC bias affects map-level variances, i.e.\ spectral-level expectations. For example, standard SpILC has the same weights as standard ILC and therefore has the same expectation bias as the ILC bias for standard ILC. To quantify the expectation bias for the constrained SpILC, we simulate 100 realizations and compare the distributions of $\hat{K}_\ell^\text{1real}$ vs $\hat{K}_\ell$ in Fig.\ \ref{fig:one_real_6noise}, and find the mean of the data-split SpILC estimators $\hat{K}^{\text{1real}}_{\ell,\text{SpILC\texttt{split}}}$ at $\ell=1500,2500,3500,4500$ to be consistent with $\hat{K}_{\ell,\text{SpILC\texttt{split}}}$ to within $0.3\%$---the weights for the latter constructed from realization-averaged spectra and thus having negligible expectation bias, with the truth well centered in the $\pm 1 \sigma$ region. This affirms that for this configuration the SpILC spectra estimator is not sensitive to expectation biases, and can safely be be measured with one sky realization without worrying about percent-level biases. 

\begin{figure*}
	\includegraphics[width=0.98\linewidth]{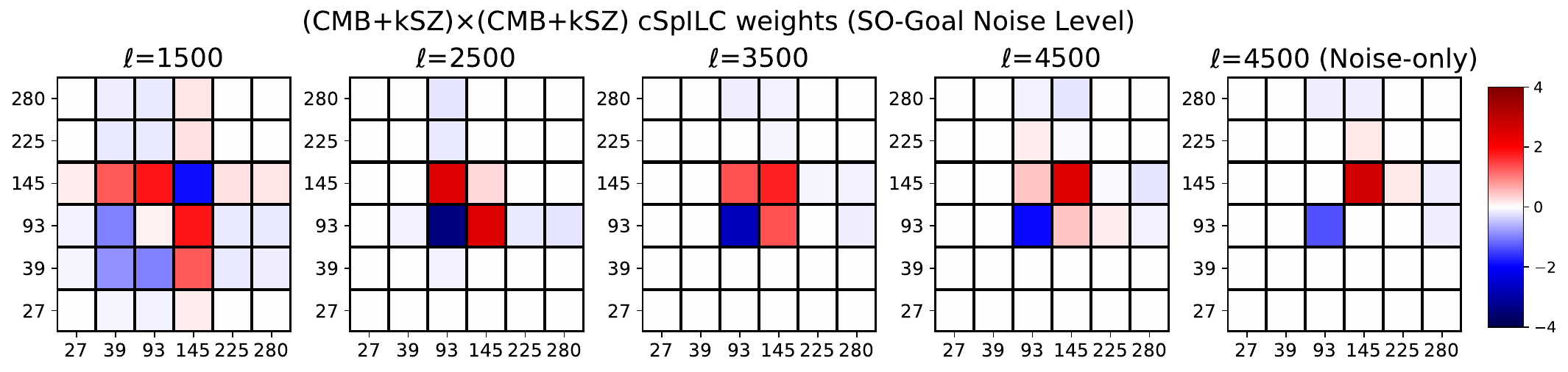}
	\caption{ Weight matrix $\hat{W}^{ij}_\ell$ visualized. The four squares from the left are the SpILC (CMB+kSZ)$\times$(CMB+kSZ) weights corresponding to the results in Fig.\ \ref{fig:temp_3comp_12noise_19real_ensemble} at $\ell=1500,2500,3500,4500$ respectively; the rightmost square is the SpILC (CMB+kSZ)$\times$(CMB+kSZ) weight at $\ell=4500$ for instrumental-noise-only input spectra $C^{ij}_\ell= N^{ij}_\ell$. The tSZ and CIB are deprojected, and the measured spectra are band-averaged with a band-width of $\Delta\ell=30$.}
	\label{fig:weight_squares_1}
	\includegraphics[width=0.98\linewidth]{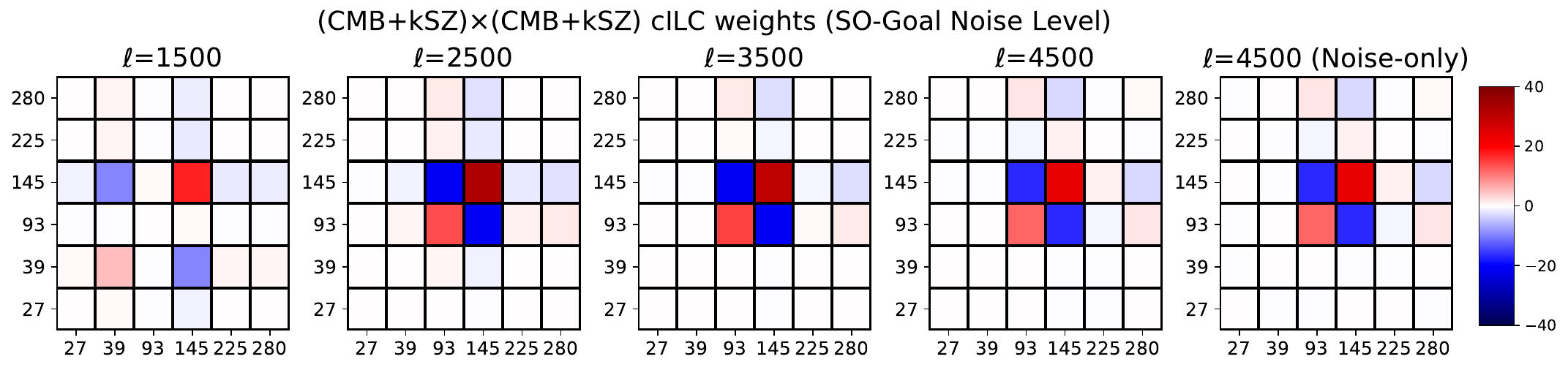}
	\caption{Similar to Fig.\ \ref{fig:weight_squares_1}, but for the cILC (CMB+kSZ)$\times$(CMB+kSZ) weights.}
	\label{fig:weight_squares_2}
\end{figure*}

\textbf{Weight matrix.} We visualize the weights $\hat{W}^{ij}_\ell$ for the (CMB+kSZ)$\times$(CMB+kSZ) spectra estimated by the non-split SpILC estimator (Fig.\ \ref{fig:weight_squares_1}) and the cILC estimator (Figs.\ \ref{fig:weight_squares_2}). Both tSZ and CIB are deprojected in both methods. We discuss two observations: 1.\ At small scales, the weights tend to the noise-only limit (noise dominates all other components in the spectra $\hat{C}^{ij}_\ell$), where they are completely determined by the ratio between noise level $\hat{N}^{ii}_\ell$ at each channel $i$, the structure of the constrained-optimization problem, and the SEDs;
2.\ the auto spectra can have negative weights, which is a novelty of SpILC; in contrary auto spectra weights for map-based ILC methods always have weights $w^i_\ell  w^i_\ell\geq 0$; 3.\ similarly, the auto and cross spectra weights for SpILC no longer have to follow  $|W^{ij}_\ell|=\sqrt{|W^{ii}_\ell||W^{jj}_\ell|}$. For example, for $\ell=4500$ in Fig.\ \ref{fig:weight_squares_1}, we have $W^{93\times 93}_{\ell=4500}=-1.92$,  $W^{145\times145}_{\ell=4500}=+2.47$, and  $W^{93\times 145}_{\ell=4500}=0.45$, therefore $|W^{93\times 145}_{\ell=4500}|<\sqrt{|W^{93\times 93}_{\ell=4500}||W^{145\times145}_{\ell=4500}|}=2.17$.

\section{Conclusion and Outlook}
\label{sec:6}
In this paper, we develop and compare spectral ILC (SpILC) against map-based constrained ILC methods using simulations comprising of CMB+kSZ+tSZ+CIB+Gaussian instrumental and atmospheric noise with SO goal noise level and beam sizes that is uncorrelated across channels. We find that both our cSpILC (CMB+kSZ)$\times$(CMB+kSZ) and tSZ$\times$tSZ estimators both achieve 7 times smaller errorbars at small scales ($\ell\gtrsim 4000$) compared to constrained ILC (tSZ+CIB or CMB+CIB deprojected), and $2$ times smaller errorbars compared to cILC with only tSZ or CMB+kSZ deprojection, respectively.

We summarize the argument why the constrained SpILC estimator can achieve lower variance than constrained ILC: we demonstrated that the constrained-minimization problem of cILC weights are equivalent to that of cSpILC with the full set of normalization and deprojection constraints in the presence of Gaussian noise. Assuming at least two components are spatially uncorrelated, we can relax one or more cSpILC constraints (which is not possible for the cILC), so that we gain more degrees of freedom to minimize the weights compared to cILC.

We further incorporate data splits in SpILC. We test our estimators with simulations. We demonstrate that our estimators indeed achieve lower variance than cILC, and that the noise bias accounts for the biases in SpILC. We find that the data-split SpILC estimators are free from noise bias and remain unbiased when the weights are obtained from a single sky realization, and only suffers a 15\% increase in standard deviation for our 2-splits implementation compared to the non-split estimators for our simulation setup.

We propose two immediate applications of Spectral ILC: the estimation of the kSZ$\times$kSZ power spectrum, as its cross-correlation with either the tSZ and CIB sums to zero; and blind component separation of the CMB polarization $BB$-spectrum, assuming it is uncorrelated with either dust or synchrotron. While the SpILC method is valid for the latter case (if SEDs are known), a needlet-space approach would be better suited than a harmonic-space approach discussed in this paper due to the anisotropy of galactic foregrounds. It will be interesting to see if cSpILC can be complementary to the needlet-space power spectrum estimator of Ref.\ \cite{kristen}.

We also envisage spectral ILC being a complementary tool for the moment ILC method \cite{cmilc}, where variations of a component SED are modeled by decomposing the component into a Taylor expansion \cite{chluba_moment} about some parameters ($(\beta,T_\text{CIB})$ for the CIB gray-body SED), and deprojecting the zeroth and higher-order terms (or \emph{moments}) in the SEDs. Moment ILC is found to be an effective mitigation to the modeling of the CIB \cite{fiona_cib}. However, given the limited number of frequency channels of current and near-future CMB experiments, the inclusion of more components through moment expansion will greatly increase the variance of ILC estimators. If spatially uncorrelated moments are available or can be devised, Spectral ILC will prove to be highly complementary by minimizing the loss of variance due to modeling of higher moments.

As high-resolution CMB experiments push millimeter-sky observations to smaller scales, the separation of the kSZ, tSZ, CIB, and other foreground emissions will be key to improving our understanding of the late-time universe. With limited number of frequency channels but more components to be modeled in order to achieve accurate component separation, we foresee Spectral ILC to play a useful role in this avenue.
\section{Acknowledgment}
JK thanks Anthony Challinor, Steven Gratton, Colin Hill, Selim Hotinli, Matthew Johnson, Terry Lam, Mathieu Remazeilles, and Kendrick Smith for discussions. We thank Mathieu Remazeilles for a careful reading of the manuscript, and Anthony Challinor for a careful reading of an earlier draft. JK acknowledges support from The Joseph Needham Foundation for Science \& Civilisation (Hong Kong) through the Joseph Needham PhD fellowship. JK is grateful for the hospitality of Perimeter Institute for Theoretical Physics, where a part of this work was done. Research at Perimeter Institute is supported in part by the Government of Canada
through the Department of Innovation, Science and Economic Development Canada and by
the Province of Ontario through the Ministry of Colleges and Universities.

This work was performed using resources provided by the Cambridge Service for Data Driven Discovery (CSD3) operated by the University of Cambridge Research Computing Service, provided by Dell EMC and Intel using Tier-2 funding from the Engineering and Physical Sciences Research Council (capital grant EP/T022159/1), and DiRAC funding from the Science and Technology Facilities Council.

\bibliography{references}

@ARTICLE{constrained_ilc,
       author = {{Remazeilles}, Mathieu and {Delabrouille}, Jacques and {Cardoso}, Jean-Fran{\c{c}}ois},
        title = "{CMB and SZ effect separation with constrained Internal Linear Combinations}",
      journal = {Monthly Notices of the Royal Astronomical Society},
         year = 2011,
        month = feb,
       volume = {410},
       number = {4},
        pages = {2481-2487},
          doi = {10.1111/j.1365-2966.2010.17624.x},
archivePrefix = {arXiv},
       eprint = {1006.5599},
 primaryClass = {astro-ph.CO},
       adsurl = {https://ui.adsabs.harvard.edu/abs/2011MNRAS.410.2481R}
}

@ARTICLE{cmilc,
       author = {{Remazeilles}, Mathieu and {Rotti}, Aditya and {Chluba}, Jens},
        title = "{Peeling off foregrounds with the constrained moment ILC method to unveil primordial CMB B modes}",
      journal = {Monthly Notices of the Royal Astronomical Society},
         year = 2021,
        month = may,
       volume = {503},
       number = {2},
        pages = {2478-2498},
          doi = {10.1093/mnras/stab648},
archivePrefix = {arXiv},
       eprint = {2006.08628},
 primaryClass = {astro-ph.CO},
       adsurl = {https://ui.adsabs.harvard.edu/abs/2021MNRAS.503.2478R}
}

@ARTICLE{hill_spergel,
       author = {{Hill}, J. Colin and {Spergel}, David N.},
        title = "{Detection of thermal SZ-CMB lensing cross-correlation in Planck nominal mission data}",
      journal = {Journal of Cosmology and Astroparticle Physics},
         year = 2014,
        month = feb,
       volume = {2014},
       number = {2},
          eid = {030},
        pages = {030},
          doi = {10.1088/1475-7516/2014/02/030},
archivePrefix = {arXiv},
       eprint = {1312.4525},
 primaryClass = {astro-ph.CO},
       adsurl = {https://ui.adsabs.harvard.edu/abs/2014JCAP...02..030H}
}

@ARTICLE{ymap_coulton,
       author = {{Coulton}, William and {Madhavacheril}, Mathew S. and {Duivenvoorden}, Adriaan J. and {Hill}, J. Colin and {Abril-Cabezas}, Irene and {Ade}, Peter A.~R. and {Aiola}, Simone and {Alford}, Tommy and {Amiri}, Mandana and {Amodeo}, Stefania and others},
        title = "{Atacama Cosmology Telescope: High-resolution component-separated maps across one third of the sky}",
      journal = {\prd},
         year = 2024,
        month = mar,
       volume = {109},
       number = {6},
          eid = {063530},
        pages = {063530},
          doi = {10.1103/PhysRevD.109.063530},
archivePrefix = {arXiv},
       eprint = {2307.01258},
 primaryClass = {astro-ph.CO},
       adsurl = {https://ui.adsabs.harvard.edu/abs/2024PhRvD.109f3530C}
}

@ARTICLE{websky,
       author = {{Stein}, George and {Alvarez}, Marcelo A. and {Bond}, J. Richard and {van Engelen}, Alexander and {Battaglia}, Nicholas},
        title = "{The Websky extragalactic CMB simulations}",
      journal = {Journal of Cosmology and Astroparticle Physics},
         year = 2020,
        month = oct,
       volume = {2020},
       number = {10},
          eid = {012},
        pages = {012},
          doi = {10.1088/1475-7516/2020/10/012},
archivePrefix = {arXiv},
       eprint = {2001.08787},
 primaryClass = {astro-ph.CO},
       adsurl = {https://ui.adsabs.harvard.edu/abs/2020JCAP...10..012S}
}

@ARTICLE{polarbear_powerspectrum,
       author = {{Ade}, P.~A.~R. and {Akiba}, Y. and {Anthony}, A.~E. and {Arnold}, K. and {Atlas}, M. and {Barron}, D. and {Boettger}, D. and {Borrill}, J. and {Chapman}, S. and {Chinone}, Y. and others},
        title = "{Measurement of the Cosmic Microwave Background Polarization Lensing Power Spectrum with the POLARBEAR Experiment}",
      journal = {\prl},
         year = 2014,
        month = jul,
       volume = {113},
       number = {2},
          eid = {021301},
        pages = {021301},
          doi = {10.1103/PhysRevLett.113.021301},
archivePrefix = {arXiv},
       eprint = {1312.6646},
 primaryClass = {astro-ph.CO},
       adsurl = {https://ui.adsabs.harvard.edu/abs/2014PhRvL.113b1301A}
}

@ARTICLE{first_kSZ,
       author = {{Hand}, Nick and {Addison}, Graeme E. and {Aubourg}, Eric and {Battaglia}, Nick and {Battistelli}, Elia S. and {Bizyaev}, Dmitry and {Bond}, J. Richard and {Brewington}, Howard and {Brinkmann}, Jon and {Brown}, Benjamin R. and others},
        title = "{Evidence of Galaxy Cluster Motions with the Kinematic Sunyaev-Zel'dovich Effect}",
      journal = {\prl},
         year = 2012,
        month = jul,
       volume = {109},
       number = {4},
          eid = {041101},
        pages = {041101},
          doi = {10.1103/PhysRevLett.109.041101},
archivePrefix = {arXiv},
       eprint = {1203.4219},
 primaryClass = {astro-ph.CO},
       adsurl = {https://ui.adsabs.harvard.edu/abs/2012PhRvL.109d1101H}
}

@ARTICLE{patchy_coulton,
       author = {{Coulton}, William R. and {Schutt}, Theo and {Maniyar}, Abhishek S. and {Schaan}, Emmanuel and {An}, Rui and {Atkins}, Zachary and {Battaglia}, Nicholas and {Bond}, J Richard and {Calabrese}, Erminia and {Choi}, Steve K. and others},
        title = "{The Atacama Cosmology Telescope: Detection of Patchy Screening of the Cosmic Microwave Background}",
      journal = {arXiv e-prints},
         year = 2024,
        month = jan,
          eid = {arXiv:2401.13033},
        pages = {arXiv:2401.13033},
          doi = {10.48550/arXiv.2401.13033},
archivePrefix = {arXiv},
       eprint = {2401.13033},
 primaryClass = {astro-ph.CO},
       adsurl = {https://ui.adsabs.harvard.edu/abs/2024arXiv240113033C}
}

@ARTICLE{delabrouille_wmap_2009,
       author = {{Delabrouille}, J. and {Cardoso}, J. -F. and {Le Jeune}, M. and {Betoule}, M. and {Fay}, G. and {Guilloux}, F.},
        title = "{A full sky, low foreground, high resolution CMB map from WMAP}",
      journal = {Astronomy and Astrophysics},
         year = 2009,
        month = jan,
       volume = {493},
       number = {3},
        pages = {835-857},
          doi = {10.1051/0004-6361:200810514},
archivePrefix = {arXiv},
       eprint = {0807.0773},
 primaryClass = {astro-ph},
       adsurl = {https://ui.adsabs.harvard.edu/abs/2009A&A...493..835D}
}

@ARTICLE{smith_lensing,
       author = {{Smith}, Kendrick M. and {Zahn}, Oliver and {Dor{\'e}}, Olivier},
        title = "{Detection of gravitational lensing in the cosmic microwave background}",
      journal = {\prd},
         year = 2007,
        month = aug,
       volume = {76},
       number = {4},
          eid = {043510},
        pages = {043510},
          doi = {10.1103/PhysRevD.76.043510},
archivePrefix = {arXiv},
       eprint = {0705.3980},
 primaryClass = {astro-ph},
       adsurl = {https://ui.adsabs.harvard.edu/abs/2007PhRvD..76d3510S}
}

@ARTICLE{planck18_comseparation,
       author = {{Planck Collaboration} and {Akrami}, Y. and {Ashdown}, M. and {Aumont}, J. and {Baccigalupi}, C. and {Ballardini}, M. and {Banday}, A.~J. and {Barreiro}, R.~B. and {Bartolo}, N. and {Basak}, S. and {Benabed}, K. and {Bersanelli}, M. and others},
        title = "{Planck 2018 results. IV. Diffuse component separation}",
      journal = {Astronomy and Astrophysics},
         year = 2020,
        month = sep,
       volume = {641},
          eid = {A4},
        pages = {A4},
          doi = {10.1051/0004-6361/201833881},
archivePrefix = {arXiv},
       eprint = {1807.06208},
 primaryClass = {astro-ph.CO},
       adsurl = {https://ui.adsabs.harvard.edu/abs/2020A&A...641A...4P}
}

@ARTICLE{tegmark_efstathiou,
       author = {{Tegmark}, Max and {Efstathiou}, George},
        title = "{A method for subtracting foregrounds from multifrequency CMB sky maps**}",
      journal = {Monthly Notices of the Royal Astronomical Society},
         year = 1996,
        month = aug,
       volume = {281},
       number = {4},
        pages = {1297-1314},
          doi = {10.1093/mnras/281.4.1297},
archivePrefix = {arXiv},
       eprint = {astro-ph/9507009},
 primaryClass = {astro-ph},
       adsurl = {https://ui.adsabs.harvard.edu/abs/1996MNRAS.281.1297T}
}

@ARTICLE{pyilc,
       author = {{McCarthy}, Fiona and {Hill}, J. Colin},
        title = "{Component-separated, CIB-cleaned thermal Sunyaev-Zel'dovich maps from Planck PR4 data with a flexible public needlet ILC pipeline}",
      journal = {\prd},
         year = 2024,
        month = jan,
       volume = {109},
       number = {2},
          eid = {023528},
        pages = {023528},
          doi = {10.1103/PhysRevD.109.023528},
archivePrefix = {arXiv},
       eprint = {2307.01043},
 primaryClass = {astro-ph.CO},
       adsurl = {https://ui.adsabs.harvard.edu/abs/2024PhRvD.109b3528M}
}

@ARTICLE{fiona_cib,
       author = {{McCarthy}, Fiona and {Hill}, J. Colin},
        title = "{Cross-correlation of the thermal Sunyaev-Zel'dovich and CMB lensing signals in Planck PR4 data with robust CIB decontamination}",
      journal = {\prd},
         year = 2024,
        month = jan,
       volume = {109},
       number = {2},
          eid = {023529},
        pages = {023529},
          doi = {10.1103/PhysRevD.109.023529},
archivePrefix = {arXiv},
       eprint = {2308.16260},
 primaryClass = {astro-ph.CO},
       adsurl = {https://ui.adsabs.harvard.edu/abs/2024PhRvD.109b3529M}
}

@ARTICLE{chluba_moment,
       author = {{Chluba}, Jens and {Hill}, James Colin and {Abitbol}, Maximilian H.},
        title = "{Rethinking CMB foregrounds: systematic extension of foreground parametrizations}",
      journal = {Monthly Notices of the Royal Astronomical Society},
         year = 2017,
        month = nov,
       volume = {472},
       number = {1},
        pages = {1195-1213},
          doi = {10.1093/mnras/stx1982},
archivePrefix = {arXiv},
       eprint = {1701.00274},
 primaryClass = {astro-ph.CO},
       adsurl = {https://ui.adsabs.harvard.edu/abs/2017MNRAS.472.1195C}
}

@ARTICLE{wmap_foreground1,
       author = {{Bennett}, C.~L. and {Hill}, R.~S. and {Hinshaw}, G. and {Nolta}, M.~R. and {Odegard}, N. and {Page}, L. and {Spergel}, D.~N. and {Weiland}, J.~L. and {Wright}, E.~L. and {Halpern}, M. and {Jarosik}, N. and {Kogut}, A. and {Limon}, M. and {Meyer}, S.~S. and {Tucker}, G.~S. and {Wollack}, E.},
        title = "{First-Year Wilkinson Microwave Anisotropy Probe (WMAP) Observations: Foreground Emission}",
      journal = {The Astrophysical Journal Supplement Series},
         year = 2003,
        month = sep,
       volume = {148},
       number = {1},
        pages = {97-117},
          doi = {10.1086/377252},
archivePrefix = {arXiv},
       eprint = {astro-ph/0302208},
 primaryClass = {astro-ph},
       adsurl = {https://ui.adsabs.harvard.edu/abs/2003ApJS..148...97B}
}

@ARTICLE{wmap_foreground2,
       author = {{Delabrouille}, J. and {Cardoso}, J. -F. and {Le Jeune}, M. and {Betoule}, M. and {Fay}, G. and {Guilloux}, F.},
        title = "{A full sky, low foreground, high resolution CMB map from WMAP}",
      journal = {Astronomy and Astrophysics},
         year = 2009,
        month = jan,
       volume = {493},
       number = {3},
        pages = {835-857},
          doi = {10.1051/0004-6361:200810514},
archivePrefix = {arXiv},
       eprint = {0807.0773},
 primaryClass = {astro-ph},
       adsurl = {https://ui.adsabs.harvard.edu/abs/2009A&A...493..835D}
}

@ARTICLE{planck_foreground13,
       author = {{Planck Collaboration} and {Ade}, P.~A.~R. and {Aghanim}, N. and {Armitage-Caplan}, C. and {Arnaud}, M. and {Ashdown}, M. and {Atrio-Barandela}, F. and {Aumont}, J. and {Baccigalupi}, C. and {Banday}, A.~J. and {Barreiro} R.~B. and others},
        title = "{Planck 2013 results. XII. Diffuse component separation}",
      journal = {Astronomy and Astrophysics},
         year = 2014,
        month = nov,
       volume = {571},
          eid = {A12},
        pages = {A12},
          doi = {10.1051/0004-6361/201321580},
archivePrefix = {arXiv},
       eprint = {1303.5072},
 primaryClass = {astro-ph.CO},
       adsurl = {https://ui.adsabs.harvard.edu/abs/2014A&A...571A..12P}
}

@ARTICLE{planck_foreground15,
       author = {{Planck Collaboration} and {Adam}, R. and {Ade}, P.~A.~R. and {Aghanim}, N. and {Arnaud}, M. and {Ashdown}, M. and {Aumont}, J. and {Baccigalupi}, C. and {Banday}, A.~J. and {Barreiro}, R.~B. and {Bartlett}, J.~G. and others},
        title = "{Planck 2015 results. IX. Diffuse component separation: CMB maps}",
      journal = {Astronomy and Astrophysics},
         year = 2016,
        month = sep,
       volume = {594},
          eid = {A9},
        pages = {A9},
          doi = {10.1051/0004-6361/201525936},
archivePrefix = {arXiv},
       eprint = {1502.05956},
 primaryClass = {astro-ph.CO},
       adsurl = {https://ui.adsabs.harvard.edu/abs/2016A&A...594A...9P}
}

@ARTICLE{act_compsep20,
       author = {{Madhavacheril}, Mathew S. and {Hill}, J. Colin and {N{\ae}ss}, Sigurd and {Addison}, Graeme E. and {Aiola}, Simone and {Baildon}, Taylor and {Battaglia}, Nicholas and {Bean}, Rachel and {Bond}, J. Richard and {Calabrese}, Erminia and {Calafut}, Victoria and {Choi}, Steve K. and {Darwish}, Omar and {Datta}, Rahul and {Devlin}, Mark J. and {Dunkley}, Joanna and {D{\"u}nner}, Rolando and {Ferraro}, Simone and {Gallardo}, Patricio A. and {Gluscevic}, Vera and {Halpern}, Mark and {Han}, Dongwon and {Hasselfield}, Matthew and {Hilton}, Matt and {Hincks}, Adam D. and {Hlo{\v{z}}ek}, Ren{\'e}e and {Ho}, Shuay-Pwu Patty and {Huffenberger}, Kevin M. and {Hughes}, John P. and {Koopman}, Brian J. and {Kosowsky}, Arthur and {Lokken}, Martine and {Louis}, Thibaut and {Lungu}, Marius and {MacInnis}, Amanda and {Maurin}, Lo{\"\i}c and {McMahon}, Jeffrey J. and {Moodley}, Kavilan and {Nati}, Federico and {Niemack}, Michael D. and {Page}, Lyman A. and {Partridge}, Bruce and {Robertson}, Naomi and {Sehgal}, Neelima and {Schaan}, Emmanuel and {Schillaci}, Alessandro and {Sherwin}, Blake D. and {Sif{\'o}n}, Crist{\'o}bal and {Simon}, Sara M. and {Spergel}, David N. and {Staggs}, Suzanne T. and {Storer}, Emilie R. and {van Engelen}, Alexander and {Vavagiakis}, Eve M. and {Wollack}, Edward J. and {Xu}, Zhilei},
        title = "{Atacama Cosmology Telescope: Component-separated maps of CMB temperature and the thermal Sunyaev-Zel'dovich effect}",
      journal = {\prd},
         year = 2020,
        month = jul,
       volume = {102},
       number = {2},
          eid = {023534},
        pages = {023534},
          doi = {10.1103/PhysRevD.102.023534},
archivePrefix = {arXiv},
       eprint = {1911.05717},
 primaryClass = {astro-ph.CO},
       adsurl = {https://ui.adsabs.harvard.edu/abs/2020PhRvD.102b3534M}
}

@ARTICLE{act_dr4_maps,
       author = {{Aiola}, Simone and {Calabrese}, Erminia and {Maurin}, Lo{\"\i}c and {Naess}, Sigurd and {Schmitt}, Benjamin L. and {Abitbol}, Maximilian H. and {Addison}, Graeme E. and {Ade}, Peter A.~R. and {Alonso}, David and {Amiri}, Mandana and others},
        title = "{The Atacama Cosmology Telescope: DR4 maps and cosmological parameters}",
      journal = {Journal of Cosmology and Astroparticle Physics},
         year = 2020,
        month = dec,
       volume = {2020},
       number = {12},
          eid = {047},
        pages = {047},
          doi = {10.1088/1475-7516/2020/12/047},
archivePrefix = {arXiv},
       eprint = {2007.07288},
 primaryClass = {astro-ph.CO},
       adsurl = {https://ui.adsabs.harvard.edu/abs/2020JCAP...12..047A}
}

@ARTICLE{act_lensing_map,
       author = {{Madhavacheril}, Mathew S. and {Qu}, Frank J. and {Sherwin}, Blake D. and {MacCrann}, Niall and {Li}, Yaqiong and {Abril-Cabezas}, Irene and {Ade}, Peter A.~R. and {Aiola}, Simone and {Alford}, Tommy and {Amiri}, Mandana and others},
        title = "{The Atacama Cosmology Telescope: DR6 Gravitational Lensing Map and Cosmological Parameters}",
      journal = {\apj},
         year = 2024,
        month = feb,
       volume = {962},
       number = {2},
          eid = {113},
        pages = {113},
          doi = {10.3847/1538-4357/acff5f},
archivePrefix = {arXiv},
       eprint = {2304.05203},
 primaryClass = {astro-ph.CO},
       adsurl = {https://ui.adsabs.harvard.edu/abs/2024ApJ...962..113M}
}

@ARTICLE{spt_release1,
       author = {{Schaffer}, K.~K. and {Crawford}, T.~M. and {Aird}, K.~A. and {Benson}, B.~A. and {Bleem}, L.~E. and {Carlstrom}, J.~E. and {Chang}, C.~L. and {Cho}, H.~M. and {Crites}, A.~T. and {de Haan}, T. and {Dobbs}, M.~A. and {George}, E.~M. and {Halverson}, N.~W. and {Holder}, G.~P. and {Holzapfel}, W.~L. and {Hoover}, S. and {Hrubes}, J.~D. and {Joy}, M. and {Keisler}, R. and {Knox}, L. and {Lee}, A.~T. and {Leitch}, E.~M. and {Lueker}, M. and {Luong-Van}, D. and {McMahon}, J.~J. and {Mehl}, J. and {Meyer}, S.~S. and {Mohr}, J.~J. and {Montroy}, T.~E. and {Padin}, S. and {Plagge}, T. and {Pryke}, C. and {Reichardt}, C.~L. and {Ruhl}, J.~E. and {Shirokoff}, E. and {Spieler}, H.~G. and {Stalder}, B. and {Staniszewski}, Z. and {Stark}, A.~A. and {Story}, K. and {Vanderlinde}, K. and {Vieira}, J.~D. and {Williamson}, R.},
        title = "{The First Public Release of South Pole Telescope Data: Maps of a 95 deg$^{2}$ Field from 2008 Observations}",
      journal = {The Astrophysical Journal},
         year = 2011,
        month = dec,
       volume = {743},
       number = {1},
          eid = {90},
        pages = {90},
          doi = {10.1088/0004-637X/743/1/90},
archivePrefix = {arXiv},
       eprint = {1111.7245},
 primaryClass = {astro-ph.CO},
       adsurl = {https://ui.adsabs.harvard.edu/abs/2011ApJ...743...90S}
}

@ARTICLE{spt_carlstrom,
       author = {{Carlstrom}, J.~E. and {Ade}, P.~A.~R. and {Aird}, K.~A. and {Benson}, B.~A. and {Bleem}, L.~E. and {Busetti}, S. and {Chang}, C.~L. and {Chauvin}, E. and {Cho}, H. -M. and {Crawford}, T.~M. and {Crites}, A.~T. and {Dobbs}, M.~A. and {Halverson}, N.~W. and {Heimsath}, S. and {Holzapfel}, W.~L. and {Hrubes}, J.~D. and {Joy}, M. and {Keisler}, R. and {Lanting}, T.~M. and {Lee}, A.~T. and {Leitch}, E.~M. and {Leong}, J. and {Lu}, W. and {Lueker}, M. and {Luong-Van}, D. and {McMahon}, J.~J. and {Mehl}, J. and {Meyer}, S.~S. and {Mohr}, J.~J. and {Montroy}, T.~E. and {Padin}, S. and {Plagge}, T. and {Pryke}, C. and {Ruhl}, J.~E. and {Schaffer}, K.~K. and {Schwan}, D. and {Shirokoff}, E. and {Spieler}, H.~G. and {Staniszewski}, Z. and {Stark}, A.~A. and {Tucker}, C. and {Vanderlinde}, K. and {Vieira}, J.~D. and {Williamson}, R.},
        title = "{The 10 Meter South Pole Telescope}",
      journal = {Publications of the Astronomical Society of the Pacific},
         year = 2011,
        month = may,
       volume = {123},
       number = {903},
        pages = {568},
          doi = {10.1086/659879},
archivePrefix = {arXiv},
       eprint = {0907.4445},
 primaryClass = {astro-ph.IM},
       adsurl = {https://ui.adsabs.harvard.edu/abs/2011PASP..123..568C}
}

@ARTICLE{sunyaev72,
       author = {{Sunyaev}, R.~A. and {Zeldovich}, Ya. B.},
        title = "{The Observations of Relic Radiation as a Test of the Nature of X-Ray Radiation from the Clusters of Galaxies}",
      journal = {Comments on Astrophysics and Space Physics},
         year = 1972,
        month = nov,
       volume = {4},
        pages = {173},
       adsurl = {https://ui.adsabs.harvard.edu/abs/1972CoASP...4..173S}
}

@ARTICLE{sunyaev80,
       author = {{Sunyaev}, R.~A. and {Zeldovich}, Ia. B.},
        title = "{Microwave background radiation as a probe of the contemporary structure and history of the universe}",
      journal = {Annual Review of Astronomy and Astrophysics},
         year = 1980,
        month = jan,
       volume = {18},
        pages = {537-560},
          doi = {10.1146/annurev.aa.18.090180.002541},
       adsurl = {https://ui.adsabs.harvard.edu/abs/1980ARA&A..18..537S}
}

@ARTICLE{planck_ymap,
       author = {{Planck Collaboration} and {Aghanim}, N. and {Arnaud}, M. and {Ashdown}, M. and {Aumont}, J. and {Baccigalupi}, C. and {Banday}, A.~J. and {Barreiro}, R.~B. and {Bartlett}, J.~G. and {Bartolo}, N. and {Battaner}, E. and others},
        title = "{Planck 2015 results. XXII. A map of the thermal Sunyaev-Zeldovich effect}",
      journal = {Astronomy and Astrophysics},
         year = 2016,
        month = sep,
       volume = {594},
          eid = {A22},
        pages = {A22},
          doi = {10.1051/0004-6361/201525826},
archivePrefix = {arXiv},
       eprint = {1502.01596},
 primaryClass = {astro-ph.CO},
       adsurl = {https://ui.adsabs.harvard.edu/abs/2016A&A...594A..22P}
}

@ARTICLE{so,
       author = {{Ade}, Peter and {Aguirre}, James and {Ahmed}, Zeeshan and {Aiola}, Simone and {Ali}, Aamir and {Alonso}, David and {Alvarez}, Marcelo A. and {Arnold}, Kam and {Ashton}, Peter and {Austermann}, Jason and {Awan}, Humna and others},
        title = "{The Simons Observatory: science goals and forecasts}",
      journal = {Journal of Cosmology and Astroparticle Physics},
         year = 2019,
        month = feb,
       volume = {2019},
       number = {2},
          eid = {056},
        pages = {056},
          doi = {10.1088/1475-7516/2019/02/056},
archivePrefix = {arXiv},
       eprint = {1808.07445},
 primaryClass = {astro-ph.CO},
       adsurl = {https://ui.adsabs.harvard.edu/abs/2019JCAP...02..056A}
}

@ARTICLE{spt_sz_catalog,
       author = {{Bleem}, L.~E. and {Stalder}, B. and {de Haan}, T. and {Aird}, K.~A. and {Allen}, S.~W. and {Applegate}, D.~E. and {Ashby}, M.~L.~N. and {Bautz}, M. and {Bayliss}, M. and {Benson}, B.~A. and others},
        title = "{Galaxy Clusters Discovered via the Sunyaev-Zel'dovich Effect in the 2500-Square-Degree SPT-SZ Survey}",
      journal = {The Astrophysical Journal Supplement Series},
         year = 2015,
        month = feb,
       volume = {216},
       number = {2},
          eid = {27},
        pages = {27},
          doi = {10.1088/0067-0049/216/2/27},
archivePrefix = {arXiv},
       eprint = {1409.0850},
 primaryClass = {astro-ph.CO},
       adsurl = {https://ui.adsabs.harvard.edu/abs/2015ApJS..216...27B}
}

@ARTICLE{act_sz_catalog,
       author = {{Hilton}, Matt and {Hasselfield}, Matthew and {Sif{\'o}n}, Crist{\'o}bal and {Battaglia}, Nicholas and {Aiola}, Simone and {Bharadwaj}, V. and {Bond}, J. Richard and {Choi}, Steve K. and {Crichton}, Devin and {Datta}, Rahul and others},
        title = "{The Atacama Cosmology Telescope: The Two-season ACTPol Sunyaev-Zel{\textquoteright}dovich Effect Selected Cluster Catalog}",
      journal = {The Astrophysical Journal Supplement Series},
         year = 2018,
        month = mar,
       volume = {235},
       number = {1},
          eid = {20},
        pages = {20},
          doi = {10.3847/1538-4365/aaa6cb},
archivePrefix = {arXiv},
       eprint = {1709.05600},
 primaryClass = {astro-ph.CO},
       adsurl = {https://ui.adsabs.harvard.edu/abs/2018ApJS..235...20H}
}

@ARTICLE{planck18_lensing,
       author = {{Planck Collaboration} and {Aghanim}, N. and {Akrami}, Y. and {Ashdown}, M. and {Aumont}, J. and {Baccigalupi}, C. and {Ballardini}, M. and {Banday}, A.~J. and {Barreiro}, R.~B. and {Bartolo}, N. and others},
        title = "{Planck 2018 results. VIII. Gravitational lensing}",
      journal = {Astronomy and Astrophysics},
         year = 2020,
        month = sep,
       volume = {641},
          eid = {A8},
        pages = {A8},
          doi = {10.1051/0004-6361/201833886},
archivePrefix = {arXiv},
       eprint = {1807.06210},
 primaryClass = {astro-ph.CO},
       adsurl = {https://ui.adsabs.harvard.edu/abs/2020A&A...641A...8P}
}

@ARTICLE{euclid,
       author = {{Laureijs}, R. and {Amiaux}, J. and {Arduini}, S. and {Augu{\`e}res}, J. -L. and {Brinchmann}, J. and {Cole}, R. and {Cropper}, M. and {Dabin}, C. and {Duvet}, L. and {Ealet}, A. and {Garilli}, B. and {Gondoin}, P. and others},
        title = "{Euclid Definition Study Report}",
      journal = {arXiv e-prints},
         year = 2011,
        month = oct,
          eid = {arXiv:1110.3193},
        pages = {arXiv:1110.3193},
          doi = {10.48550/arXiv.1110.3193},
archivePrefix = {arXiv},
       eprint = {1110.3193},
 primaryClass = {astro-ph.CO},
       adsurl = {https://ui.adsabs.harvard.edu/abs/2011arXiv1110.3193L}
}

@ARTICLE{lsst,
       author = {{LSST Science Collaboration} and {Abell}, Paul A. and {Allison}, Julius and {Anderson}, Scott F. and {Andrew}, John R. and {Angel}, J. Roger P. and {Armus}, Lee and {Arnett}, David and {Asztalos}, S.~J. and {Axelrod}, Tim S. and {Bailey}, Stephen and others},
        title = "{LSST Science Book, Version 2.0}",
      journal = {arXiv e-prints},
         year = 2009,
        month = dec,
          eid = {arXiv:0912.0201},
        pages = {arXiv:0912.0201},
          doi = {10.48550/arXiv.0912.0201},
archivePrefix = {arXiv},
       eprint = {0912.0201},
 primaryClass = {astro-ph.IM},
       adsurl = {https://ui.adsabs.harvard.edu/abs/2009arXiv0912.0201L}
}

@ARTICLE{desi,
       author = {{DESI Collaboration} and {Aghamousa}, Amir and {Aguilar}, Jessica and {Ahlen}, Steve and {Alam}, Shadab and {Allen}, Lori E. and {Allende Prieto}, Carlos and {Annis}, James and {Bailey}, Stephen and {Balland}, Christophe and {Ballester}, Otger and others},
        title = "{The DESI Experiment Part I: Science,Targeting, and Survey Design}",
      journal = {arXiv e-prints},
         year = 2016,
        month = oct,
          eid = {arXiv:1611.00036},
        pages = {arXiv:1611.00036},
          doi = {10.48550/arXiv.1611.00036},
archivePrefix = {arXiv},
       eprint = {1611.00036},
 primaryClass = {astro-ph.IM},
       adsurl = {https://ui.adsabs.harvard.edu/abs/2016arXiv161100036D}
}

@ARTICLE{des,
       author = {{The Dark Energy Survey Collaboration}},
        title = "{The Dark Energy Survey}",
      journal = {arXiv e-prints},
         year = 2005,
        month = oct,
          eid = {astro-ph/0510346},
        pages = {astro-ph/0510346},
          doi = {10.48550/arXiv.astro-ph/0510346},
archivePrefix = {arXiv},
       eprint = {astro-ph/0510346},
 primaryClass = {astro-ph},
       adsurl = {https://ui.adsabs.harvard.edu/abs/2005astro.ph.10346T}
}

@ARTICLE{eriksen06,
       author = {{Eriksen}, H.~K. and {Dickinson}, C. and {Lawrence}, C.~R. and {Baccigalupi}, C. and {Banday}, A.~J. and {G{\'o}rski}, K.~M. and {Hansen}, F.~K. and {Lilje}, P.~B. and {Pierpaoli}, E. and {Seiffert}, M.~D. and {Smith}, K.~M. and {Vanderlinde}, K.},
        title = "{Cosmic Microwave Background Component Separation by Parameter Estimation}",
      journal = {The Astrophysical Journal},
         year = 2006,
        month = apr,
       volume = {641},
       number = {2},
        pages = {665-682},
          doi = {10.1086/500499},
archivePrefix = {arXiv},
       eprint = {astro-ph/0508268},
 primaryClass = {astro-ph},
       adsurl = {https://ui.adsabs.harvard.edu/abs/2006ApJ...641..665E}
}

@ARTICLE{eriksen08,
       author = {{Eriksen}, H.~K. and {Jewell}, J.~B. and {Dickinson}, C. and {Banday}, A.~J. and {G{\'o}rski}, K.~M. and {Lawrence}, C.~R.},
        title = "{Joint Bayesian Component Separation and CMB Power Spectrum Estimation}",
      journal = {The Astrophysical Journal},
         year = 2008,
        month = mar,
       volume = {676},
       number = {1},
        pages = {10-32},
          doi = {10.1086/525277},
archivePrefix = {arXiv},
       eprint = {0709.1058},
 primaryClass = {astro-ph},
       adsurl = {https://ui.adsabs.harvard.edu/abs/2008ApJ...676...10E}
}

@ARTICLE{planck15_diffuse,
       author = {{Planck Collaboration} and {Adam}, R. and {Ade}, P.~A.~R. and {Aghanim}, N. and {Alves}, M.~I.~R. and {Arnaud}, M. and {Ashdown}, M. and {Aumont}, J. and {Baccigalupi}, C. and {Banday}, A.~J. and {Barreiro}, R.~B. and others},
        title = "{Planck 2015 results. X. Diffuse component separation: Foreground maps}",
      journal = {Astronomy and Astrophysics},
         year = 2016,
        month = sep,
       volume = {594},
          eid = {A10},
        pages = {A10},
          doi = {10.1051/0004-6361/201525967},
archivePrefix = {arXiv},
       eprint = {1502.01588},
 primaryClass = {astro-ph.CO},
       adsurl = {https://ui.adsabs.harvard.edu/abs/2016A&A...594A..10P}
}

@ARTICLE{sevem03,
       author = {{Mart{\'\i}nez-Gonz{\'a}lez}, E. and {Diego}, J.~M. and {Vielva}, P. and {Silk}, J.},
        title = "{Cosmic microwave background power spectrum estimation and map reconstruction with the expectation-maximization algorithm}",
      journal = {Monthly Notices of the Royal Astronomical Society},
         year = 2003,
        month = nov,
       volume = {345},
       number = {4},
        pages = {1101-1109},
          doi = {10.1046/j.1365-2966.2003.06885.x},
archivePrefix = {arXiv},
       eprint = {astro-ph/0302094},
 primaryClass = {astro-ph},
       adsurl = {https://ui.adsabs.harvard.edu/abs/2003MNRAS.345.1101M}
}

@ARTICLE{sevem08,
       author = {{Leach}, S.~M. and {Cardoso}, J. -F. and {Baccigalupi}, C. and {Barreiro}, R.~B. and {Betoule}, M. and {Bobin}, J. and {Bonaldi}, A. and {Delabrouille}, J. and {de Zotti}, G. and {Dickinson}, C. and {Eriksen}, H.~K. and {Gonz{\'a}lez-Nuevo}, J. and {Hansen}, F.~K. and {Herranz}, D. and {Le Jeune}, M. and {L{\'o}pez-Caniego}, M. and {Mart{\'\i}nez-Gonz{\'a}lez}, E. and {Massardi}, M. and {Melin}, J. -B. and {Miville-Desch{\^e}nes}, M. -A. and {Patanchon}, G. and {Prunet}, S. and {Ricciardi}, S. and {Salerno}, E. and {Sanz}, J.~L. and {Starck}, J. -L. and {Stivoli}, F. and {Stolyarov}, V. and {Stompor}, R. and {Vielva}, P.},
        title = "{Component separation methods for the PLANCK mission}",
      journal = {Astronomy and Astrophysics},
         year = 2008,
        month = nov,
       volume = {491},
       number = {2},
        pages = {597-615},
          doi = {10.1051/0004-6361:200810116},
archivePrefix = {arXiv},
       eprint = {0805.0269},
 primaryClass = {astro-ph},
       adsurl = {https://ui.adsabs.harvard.edu/abs/2008A&A...491..597L}
}

@ARTICLE{sevem12,
       author = {{Fern{\'a}ndez-Cobos}, R. and {Vielva}, P. and {Barreiro}, R.~B. and {Mart{\'\i}nez-Gonz{\'a}lez}, E.},
        title = "{Multiresolution internal template cleaning: an application to the Wilkinson Microwave Anisotropy Probe 7-yr polarization data}",
      journal = {Monthly Notices of the Royal Astronomical Society},
         year = 2012,
        month = mar,
       volume = {420},
       number = {3},
        pages = {2162-2169},
          doi = {10.1111/j.1365-2966.2011.20182.x},
archivePrefix = {arXiv},
       eprint = {1106.2016},
 primaryClass = {astro-ph.CO},
       adsurl = {https://ui.adsabs.harvard.edu/abs/2012MNRAS.420.2162F}
}

@ARTICLE{eriksen_ilc,
       author = {{Eriksen}, H.~K. and {Banday}, A.~J. and {G{\'o}rski}, K.~M. and {Lilje}, P.~B.},
        title = "{On Foreground Removal from the Wilkinson Microwave Anisotropy Probe Data by an Internal Linear Combination Method: Limitations and Implications}",
      journal = {The Astrophysical Journal},
         year = 2004,
        month = sep,
       volume = {612},
       number = {2},
        pages = {633-646},
          doi = {10.1086/422807},
archivePrefix = {arXiv},
       eprint = {astro-ph/0403098},
 primaryClass = {astro-ph},
       adsurl = {https://ui.adsabs.harvard.edu/abs/2004ApJ...612..633E}
}

@ARTICLE{tegmark_ilc,
       author = {{Tegmark}, Max and {de Oliveira-Costa}, Ang{\'e}lica and {Hamilton}, Andrew J.},
        title = "{High resolution foreground cleaned CMB map from WMAP}",
      journal = {\prd},
         year = 2003,
        month = dec,
       volume = {68},
       number = {12},
          eid = {123523},
        pages = {123523},
          doi = {10.1103/PhysRevD.68.123523},
archivePrefix = {arXiv},
       eprint = {astro-ph/0302496},
 primaryClass = {astro-ph},
       adsurl = {https://ui.adsabs.harvard.edu/abs/2003PhRvD..68l3523T}
}

@ARTICLE{cilc,
       author = {{Remazeilles}, Mathieu and {Delabrouille}, Jacques and {Cardoso}, Jean-Fran{\c{c}}ois},
        title = "{CMB and SZ effect separation with constrained Internal Linear Combinations}",
      journal = {Monthly Notices of the Royal Astronomical Society},
         year = 2011,
        month = feb,
       volume = {410},
       number = {4},
        pages = {2481-2487},
          doi = {10.1111/j.1365-2966.2010.17624.x},
archivePrefix = {arXiv},
       eprint = {1006.5599},
 primaryClass = {astro-ph.CO},
       adsurl = {https://ui.adsabs.harvard.edu/abs/2011MNRAS.410.2481R}
}

@ARTICLE{kristen,
       author = {{Surrao}, Kristen M. and {Hill}, J. Colin},
        title = "{Constraining cosmological parameters with needlet internal linear combination maps. I. Analytic power spectrum formalism}",
      journal = {\prd},
         year = 2024,
        month = sep,
       volume = {110},
       number = {6},
          eid = {063509},
        pages = {063509},
          doi = {10.1103/PhysRevD.110.063509},
archivePrefix = {arXiv},
       eprint = {2403.02261},
 primaryClass = {astro-ph.CO},
       adsurl = {https://ui.adsabs.harvard.edu/abs/2024PhRvD.110f3509S}
}

@ARTICLE{choi2020,
       author = {{Choi}, Steve K. and {Hasselfield}, Matthew and {Ho}, Shuay-Pwu Patty and {Koopman}, Brian and {Lungu}, Marius and {Abitbol}, Maximilian H. and {Addison}, Graeme E. and {Ade}, Peter A.~R. and {Aiola}, Simone and {Alonso}, David and others},
        title = "{The Atacama Cosmology Telescope: a measurement of the Cosmic Microwave Background power spectra at 98 and 150 GHz}",
      journal = {Journal of Cosmology and Astroparticle Physics},
         year = 2020,
        month = dec,
       volume = {2020},
       number = {12},
          eid = {045},
        pages = {045},
          doi = {10.1088/1475-7516/2020/12/045},
archivePrefix = {arXiv},
       eprint = {2007.07289},
 primaryClass = {astro-ph.CO},
       adsurl = {https://ui.adsabs.harvard.edu/abs/2020JCAP...12..045C}
}

@ARTICLE{maccrann23,
       author = {{MacCrann}, Niall and {Sherwin}, Blake D. and {Qu}, Frank J. and {Namikawa}, Toshiya and {Madhavacheril}, Mathew S. and {Abril-Cabezas}, Irene and {An}, Rui and {Austermann}, Jason E. and {Battaglia}, Nicholas and {Battistelli}, Elia S. and {Beall}, James A. and {Bolliet}, Boris and {Bond}, J. Richard and {Cai}, Hongbo and {Calabrese}, Erminia and {Coulton}, William R. and {Darwish}, Omar and {Duff}, Shannon M. and {Duivenvoorden}, Adriaan J. and {Dunkley}, Jo and {Farren}, Gerrit S. and {Ferraro}, Simone and {Golec}, Joseph E. and {Guan}, Yilun and {Han}, Dongwon and {Herv{\'\i}as-Caimapo}, Carlos and {Hill}, J. Colin and {Hilton}, Matt and {Hlo{\v{z}}ek}, Ren{\'e}e and {Hubmayr}, Johannes and {Kim}, Joshua and {Li}, Zack and {Kosowsky}, Arthur and {Louis}, Thibaut and {McMahon}, Jeff and {Marques}, Gabriela A. and {Moodley}, Kavilan and {Naess}, Sigurd and {Niemack}, Michael D. and {Page}, Lyman and {Partridge}, Bruce and {Schaan}, Emmanuel and {Sehgal}, Neelima and {Sif{\'o}n}, Crist{\'o}bal and {Wollack}, Edward J. and {Salatino}, Maria and {Ullom}, Joel N. and {Van Lanen}, Jeff and {Van Engelen}, Alexander and {Wenzl}, Lukas},
        title = "{The Atacama Cosmology Telescope: Mitigating the Impact of Extragalactic Foregrounds for the DR6 Cosmic Microwave Background Lensing Analysis}",
      journal = {\apj},
         year = 2024,
        month = may,
       volume = {966},
       number = {1},
          eid = {138},
        pages = {138},
          doi = {10.3847/1538-4357/ad2610},
archivePrefix = {arXiv},
       eprint = {2304.05196},
 primaryClass = {astro-ph.CO},
       adsurl = {https://ui.adsabs.harvard.edu/abs/2024ApJ...966..138M}
}

@ARTICLE{2020A&A...641A...1P,
       author = {{Planck Collaboration} and {Aghanim}, N. and {Akrami}, Y. and {Arroja}, F. and {Ashdown}, M. and {Aumont}, J. and {Baccigalupi}, C. and {Ballardini}, M. and {Banday}, A.~J. and {Barreiro}, R.~B. and {Bartolo}, N. and others},
        title = "{Planck 2018 results. I. Overview and the cosmological legacy of Planck}",
      journal = {Astronomy and Astrophysics},
         year = 2020,
        month = sep,
       volume = {641},
          eid = {A1},
        pages = {A1},
          doi = {10.1051/0004-6361/201833880},
archivePrefix = {arXiv},
       eprint = {1807.06205},
 primaryClass = {astro-ph.CO},
       adsurl = {https://ui.adsabs.harvard.edu/abs/2020A&A...641A...1P}
}

@ARTICLE{2015arXiv151002809H,
       author = {{Henderson}, S.~W. and {Allison}, R. and {Austermann}, J. and {Baildon}, T. and {Battaglia}, N. and {Beall}, J.~A. and {Becker}, D. and {De Bernardis}, F. and {Bond}, J.~R. and {Calabrese}, E. and others},
        title = "{Advanced ACTPol Cryogenic Detector Arrays and Readout}",
      journal = {arXiv e-prints},
         year = 2015,
        month = oct,
          eid = {arXiv:1510.02809},
        pages = {arXiv:1510.02809},
          doi = {10.48550/arXiv.1510.02809},
archivePrefix = {arXiv},
       eprint = {1510.02809},
 primaryClass = {astro-ph.IM},
       adsurl = {https://ui.adsabs.harvard.edu/abs/2015arXiv151002809H}
}

@INPROCEEDINGS{2014SPIE.9153E..1PB,
       author = {{Benson}, B.~A. and {Ade}, P.~A.~R. and {Ahmed}, Z. and {Allen}, S.~W. and {Arnold}, K. and {Austermann}, J.~E. and {Bender}, A.~N. and {Bleem}, L.~E. and {Carlstrom}, J.~E. and {Chang}, C.~L. and others},
        title = "{SPT-3G: a next-generation cosmic microwave background polarization experiment on the South Pole telescope}",
    booktitle = {Millimeter, Submillimeter, and Far-Infrared Detectors and Instrumentation for Astronomy VII},
         year = 2014,
       editor = {{Holland}, Wayne S. and {Zmuidzinas}, Jonas},
       series = {Society of Photo-Optical Instrumentation Engineers (SPIE) Conference Series},
       volume = {9153},
        month = jul,
          eid = {91531P},
        pages = {91531P},
          doi = {10.1117/12.2057305},
archivePrefix = {arXiv},
       eprint = {1407.2973},
 primaryClass = {astro-ph.IM},
       adsurl = {https://ui.adsabs.harvard.edu/abs/2014SPIE.9153E..1PB}
}

@INPROCEEDINGS{spherex,
       author = {{Crill}, Brendan P. and {Werner}, Michael and {Akeson}, Rachel and {Ashby}, Matthew and {Bleem}, Lindsey and {Bock}, James J. and {Bryan}, Sean and {Burnham}, Jill and {Byunh}, Joyce and {Chang}, Tzu-Ching and others},
        title = "{SPHEREx: NASA's near-infrared spectrophotometric all-sky survey}",
    booktitle = {Space Telescopes and Instrumentation 2020: Optical, Infrared, and Millimeter Wave},
         year = 2020,
       editor = {{Lystrup}, Makenzie and {Perrin}, Marshall D.},
       series = {Society of Photo-Optical Instrumentation Engineers (SPIE) Conference Series},
       volume = {11443},
        month = dec,
          eid = {114430I},
        pages = {114430I},
          doi = {10.1117/12.2567224},
archivePrefix = {arXiv},
       eprint = {2404.11017},
 primaryClass = {astro-ph.IM},
       adsurl = {https://ui.adsabs.harvard.edu/abs/2020SPIE11443E..0IC}
}

@ARTICLE{spt_lensing23,
       author = {{Omori}, Y. and {Baxter}, E.~J. and {Chang}, C. and {Friedrich}, O. and {Alarcon}, A. and {Alves}, O. and {Amon}, A. and {Andrade-Oliveira}, F. and {Bechtol}, K. and {Becker}, M.~R. and others},
        title = "{Joint analysis of Dark Energy Survey Year 3 data and CMB lensing from SPT and Planck. I. Construction of CMB lensing maps and modeling choices}",
      journal = {\prd},
         year = 2023,
        month = jan,
       volume = {107},
       number = {2},
          eid = {023529},
        pages = {023529},
          doi = {10.1103/PhysRevD.107.023529},
archivePrefix = {arXiv},
       eprint = {2203.12439},
 primaryClass = {astro-ph.CO},
       adsurl = {https://ui.adsabs.harvard.edu/abs/2023PhRvD.107b3529O}
}

@ARTICLE{dvorkin09,
       author = {{Dvorkin}, Cora and {Smith}, Kendrick M.},
        title = "{Reconstructing patchy reionization from the cosmic microwave background}",
      journal = {\prd},
         year = 2009,
        month = feb,
       volume = {79},
       number = {4},
          eid = {043003},
        pages = {043003},
          doi = {10.1103/PhysRevD.79.043003},
archivePrefix = {arXiv},
       eprint = {0812.1566},
 primaryClass = {astro-ph},
       adsurl = {https://ui.adsabs.harvard.edu/abs/2009PhRvD..79d3003D}
}

@ARTICLE{spt_ksz21,
       author = {{Reichardt}, C.~L. and {Patil}, S. and {Ade}, P.~A.~R. and {Anderson}, A.~J. and {Austermann}, J.~E. and {Avva}, J.~S. and {Baxter}, E. and {Beall}, J.~A. and {Bender}, A.~N. and {Benson}, B.~A. and others},
        title = "{An Improved Measurement of the Secondary Cosmic Microwave Background Anisotropies from the SPT-SZ + SPTpol Surveys}",
      journal = {The Astrophysical Journal},
         year = 2021,
        month = feb,
       volume = {908},
       number = {2},
          eid = {199},
        pages = {199},
          doi = {10.3847/1538-4357/abd407},
archivePrefix = {arXiv},
       eprint = {2002.06197},
 primaryClass = {astro-ph.CO},
       adsurl = {https://ui.adsabs.harvard.edu/abs/2021ApJ...908..199R}
}

@ARTICLE{schaan20,
       author = {{Schaan}, Emmanuel and {Ferraro}, Simone and {Amodeo}, Stefania and {Battaglia}, Nicholas and {Aiola}, Simone and {Austermann}, Jason E. and {Beall}, James A. and {Bean}, Rachel and {Becker}, Daniel T. and {Bond}, Richard J. and others},
        title = "{Atacama Cosmology Telescope: Combined kinematic and thermal Sunyaev-Zel'dovich measurements from BOSS CMASS and LOWZ halos}",
      journal = {\prd},
         year = 2021,
        month = mar,
       volume = {103},
       number = {6},
          eid = {063513},
        pages = {063513},
          doi = {10.1103/PhysRevD.103.063513},
archivePrefix = {arXiv},
       eprint = {2009.05557},
 primaryClass = {astro-ph.CO},
       adsurl = {https://ui.adsabs.harvard.edu/abs/2021PhRvD.103f3513S}
}

@article{smith_ferraro,
  title = {Detecting Patchy Reionization in the Cosmic Microwave Background},
  author = {Smith, Kendrick M. and Ferraro, Simone},
  journal = {Phys. Rev. Lett.},
  volume = {119},
  issue = {2},
  pages = {021301},
  numpages = {5},
  year = {2017},
  month = {Jul},
  publisher = {American Physical Society},
  doi = {10.1103/PhysRevLett.119.021301},
  url = {https://link.aps.org/doi/10.1103/PhysRevLett.119.021301}
}

@ARTICLE{kompaneets,
       author = {{Kompaneets}, A.~S.},
        title = "{The Establishment of Thermal Equilibrium between Quanta and Electrons}",
      journal = {Soviet Journal of Experimental and Theoretical Physics},
         year = 1957,
        month = may,
       volume = {4},
       number = {5},
        pages = {730-737},
       adsurl = {https://ui.adsabs.harvard.edu/abs/1957JETP....4..730K}
}

@ARTICLE{chen09,
       author = {{Chen}, X. and {Wright}, E.~L.},
        title = "{Extragalactic Point Source Search in Five-Year WMAP 41, 61, and 94 Ghz Maps}",
      journal = {\apj},
         year = 2009,
        month = mar,
       volume = {694},
       number = {1},
        pages = {222-234},
          doi = {10.1088/0004-637X/694/1/222},
archivePrefix = {arXiv},
       eprint = {0809.4025},
 primaryClass = {astro-ph},
       adsurl = {https://ui.adsabs.harvard.edu/abs/2009ApJ...694..222C}
}

@ARTICLE{spt_4pksz,
       author = {{Raghunathan}, S. and {Ade}, P.~A.~R. and {Anderson}, A.~J. and {Ansarinejad}, B. and {Archipley}, M. and {Austermann}, J.~E. and {Balkenhol}, L. and {Beall}, J.~A. and {Benabed}, K. and {Bender}, A.~N. and {Benson}, B.~A. and {Bianchini}, F. and {Bleem}, L.~E. and {Bock}, J. and {Bouchet}, F.~R. and {Bryant}, L. and {Camphuis}, E. and {Carlstrom}, J.~E. and {Cecil}, T.~W. and {Chang}, C.~L. and {Chaubal}, P. and {Chiang}, H.~C. and {Chichura}, P.~M. and {Chou}, T. -L. and {Citron}, R. and {Coerver}, A. and {Crawford}, T.~M. and {Crites}, A.~T. and {Cukierman}, A. and {Daley}, C. and {Dibert}, K.~R. and {Dobbs}, M.~A. and {Doussot}, A. and {Dutcher}, D. and {Everett}, W. and {Feng}, C. and {Ferguson}, K.~R. and {Fichman}, K. and {Foster}, A. and {Galli}, S. and {Gallicchio}, J. and {Gambrel}, A.~E. and {Gardner}, R.~W. and {Ge}, F. and {George}, E.~M. and {Goeckner-Wald}, N. and {Gualtieri}, R. and {Guidi}, F. and {Guns}, S. and {Gupta}, N. and {de Haan}, T. and {Halverson}, N.~W. and {Hivon}, E. and {Holder}, G.~P. and {Holzapfel}, W.~L. and {Hood}, J.~C. and {Hrubes}, J.~D. and {Hryciuk}, A. and {Huang}, N. and {Hubmayr}, J. and {Irwin}, K.~D. and {K{\'e}ruzor{\'e}}, F. and {Khalife}, A.~R. and {Knox}, L. and {Korman}, M. and {Kornoelje}, K. and {Kuo}, C. -L. and {Lee}, A.~T. and {Levy}, K. and {Li}, D. and {Lowitz}, A.~E. and {Lu}, C. and {Maniyar}, A. and {Martsen}, E.~S. and {McMahon}, J.~J. and {Menanteau}, F. and {Millea}, M. and {Montgomery}, J. and {Corbett Moran}, C. and {Nakato}, Y. and {Natoli}, T. and {Nibarger}, J.~P. and {Noble}, G.~I. and {Novosad}, V. and {Omori}, Y. and {Padin}, S. and {Pan}, Z. and {Paschos}, P. and {Patil}, S. and {Phadke}, K.~A. and {Prabhu}, K. and {Pryke}, C. and {Quan}, W. and {Rahimi}, M. and {Rahlin}, A. and {Reichardt}, C.~L. and {Rouble}, M. and {Ruhl}, J.~E. and {Saliwanchik}, B.~R. and {Schaffer}, K.~K. and {Schiappucci}, E. and {Sievers}, C. and {Smecher}, G. and {Sobrin}, J.~A. and {Stark}, A.~A. and {Stephen}, J. and {Suzuki}, A. and {Tandoi}, C. and {Thompson}, K.~L. and {Thorne}, B. and {Trendafilova}, C. and {Tucker}, C. and {Umilta}, C. and {Veach}, T. and {Vieira}, J.~D. and {Viero}, M.~P. and {Wan}, Y. and {Wang}, G. and {Whitehorn}, N. and {Wu}, W.~L.~K. and {Yefremenko}, V. and {Young}, M.~R. and {Zebrowski}, J.~A. and {Zemcov}, M. and {SPT-3G} and {SPTpol Collaboration}},
        title = "{First Constraints on the Epoch of Reionization Using the Non-Gaussianity of the Kinematic Sunyaev-Zel'dovich Effect from the South Pole Telescope and Herschel-SPIRE Observations}",
      journal = {\prl},
         year = 2024,
        month = sep,
       volume = {133},
       number = {12},
          eid = {121004},
        pages = {121004},
          doi = {10.1103/PhysRevLett.133.121004},
archivePrefix = {arXiv},
       eprint = {2403.02337},
 primaryClass = {astro-ph.CO},
       adsurl = {https://ui.adsabs.harvard.edu/abs/2024PhRvL.133l1004R}
}

@ARTICLE{maccrann_4pksz,
       author = {{MacCrann}, Niall and {Qu}, Frank J. and {Namikawa}, Toshiya and {Bolliet}, Boris and {Cai}, Hongbo and {Calabrese}, Erminia and {Choi}, Steve K. and {Coulton}, William and {Darwish}, Omar and {Ferraro}, Simone and {Guan}, Yilun and {Hill}, J. Colin and {Hilton}, Matt and {Hlo{\v{z}}ek}, Ren{\'e}e and {Kramer}, Darby and {Madhavacheril}, Mathew S. and {Moodley}, Kavilan and {Sehgal}, Neelima and {Sherwin}, Blake D. and {Sif{\'o}n}, Crist{\'o}bal and {Staggs}, Suzanne T. and {Trac}, Hy and {Van Engelen}, Alexander and {Vavagiakis}, Eve M.},
        title = "{The Atacama Cosmology Telescope: Reionization kSZ trispectrum methodology and limits}",
      journal = {Monthly Notices of the Royal Astronomical Society},
         year = 2024,
        month = aug,
       volume = {532},
       number = {4},
        pages = {4247-4260},
          doi = {10.1093/mnras/stae1746},
archivePrefix = {arXiv},
       eprint = {2405.01188},
 primaryClass = {astro-ph.CO},
       adsurl = {https://ui.adsabs.harvard.edu/abs/2024MNRAS.532.4247M}
}

@misc{so_noise_model,
  author = {Hasselfield, Matthew and Namikawa, Toshiya and  Zonca, Andrea and Madhavacheril, Mathew S. and Hill, Colin and others},
  title = {so\_noise\_models},
  year = {2022},
  publisher = {GitHub},
  journal = {GitHub repository},
  howpublished = {\url{https://github.com/simonsobs/so_noise_models}}}

@ARTICLE{irene25,
       author = {{Abril-Cabezas}, Irene and {Qu}, Frank J. and {Sherwin}, Blake D. and {van Engelen}, Alexander and {MacCrann}, Niall and {Herv{\'\i}as-Caimapo}, Carlos and {Darwish}, Omar and {Hill}, J. Colin and {Madhavacheril}, Mathew S. and {Sehgal}, Neelima},
        title = "{Impact of Galactic non-Gaussian foregrounds on CMB lensing measurements}",
      journal = {arXiv e-prints},
         year = 2025,
        month = may,
          eid = {arXiv:2505.03737},
        pages = {arXiv:2505.03737},
          doi = {10.48550/arXiv.2505.03737},
archivePrefix = {arXiv},
       eprint = {2505.03737},
 primaryClass = {astro-ph.CO},
       adsurl = {https://ui.adsabs.harvard.edu/abs/2025arXiv250503737A}
}

@ARTICLE{abs1,
       author = {{Zhang}, Pengjie and {Zhang}, Jun and {Zhang}, Le},
        title = "{ABS: an analytical method of blind separation of CMB from foregrounds}",
      journal = {Monthly Notices of the Royal Astronomical Society},
         year = 2019,
        month = apr,
       volume = {484},
       number = {2},
        pages = {1616-1626},
          doi = {10.1093/mnras/stz091},
       adsurl = {https://ui.adsabs.harvard.edu/abs/2019MNRAS.484.1616Z}
}

@ARTICLE{abs2,
       author = {{Yao}, Jian and {Zhang}, Le and {Zhao}, Yuxi and {Zhang}, Pengjie and {Santos}, Larissa and {Zhang}, Jun},
        title = "{Testing the ABS Method with the Simulated Planck Temperature Maps}",
      journal = {The Astrophysical Journal Supplement Series},
         year = 2018,
        month = dec,
       volume = {239},
       number = {2},
          eid = {36},
        pages = {36},
          doi = {10.3847/1538-4365/aaef7a},
archivePrefix = {arXiv},
       eprint = {1807.07016},
 primaryClass = {astro-ph.CO},
       adsurl = {https://ui.adsabs.harvard.edu/abs/2018ApJS..239...36Y}
}

\appendix

\section{Detailed derivations of SpILC weights}
\label{app:a}
\subsection{Derivation for SpILC weights}
\label{app:derivation_spilc}
Our objective is to find the weights $W^{ij}_\ell$ in the constrained-optimization problem
 \begin{align}
	\begin{cases}
		\partial_{W^{ij}_\ell} \left[W^{ab}_\ell W^{cd}_\ell C^{ac}_\ell C^{bd}_\ell - \lambda(W^{cd}_\ell a^ca^d-1) \right] =0 & \mbox{, $i\leq j$}\\
		W^{cd}_\ell a^c a^d = 1 \; ,
	\end{cases} \; .
	\label{eq:pilc_eqs_app}
\end{align}
Noting that $W_{\ell}^{ij}$ is symmetric,
\begin{align}
	\sum_c^N \sum_d^N W_{\ell}^{cd} &= W_{\ell}^{11} + W_{\ell}^{12}+ \cdots + W_{\ell}^{21} + W_{\ell}^{22} + \cdots \nonumber \\
				 &= W_{\ell}^{11} + 2W_{\ell}^{12}+ \cdots + 2W_{\ell}^{1N} + W_{\ell}^{22} + 2W_{\ell}^{23}+\cdots \nonumber \\
				 &= \sum_{c}^N \sum_d^c (2-\delta_{cd})W_{\ell}^{cd} \equiv \sum^{c \leq d}_{c,d} (2-\delta_{cd}) W_{\ell}^{cd} \; .
\end{align}
Defining
\begin{align}
	\bar{W}_{cd} \equiv (2-\delta_{cd})W_{\ell}^{cd} = 
	\begin{cases}
		W_{\ell}^{cd} &\mbox{for $c=d$},\\
		2W_{\ell}^{cd} &\mbox{for $c\neq d$,}
	\end{cases}
\end{align}
We can rewrite sums over frequency channels to sums over the $M$ degrees of freedom of our weights:
\begin{align}
	\sum_{c,d}^N W_{\ell}^{cd}C_{\ell}^{ac}C_{\ell}^{bd} &= \sum^{c\leq d}_{c,d} (2-\delta_{cd}) W_{\ell}^{cd} \frac{(C_{\ell}^{ac} C_{\ell}^{bd}+ C_{\ell}^{ad} C_{\ell}^{bc})}{2} \nonumber \\
						    &\equiv \sum^{c\leq d}_{c,d} \bar{W}_\ell^{cd} C_{\ell}^{a(c|} C_{\ell}^{b|d)} \; ,
						    \label{eq:sum_manipulation}
\end{align}
where $C_{\ell}^{a(c|}C_{\ell}^{b|d)}\equiv (C_{\ell}^{ac} C_{\ell}^{bd}+ C_{\ell}^{ad} C_{\ell}^{bc})/2$.

We rewrite the set of simultaneous equations in Eq.\ (\ref{eq:pilc_eqs}) in terms of sums over the weight degrees of freedom:
 \begin{align}
	\implies &\begin{cases}
		\partial_{\bar{W}_\ell^{ij}}  \left[ \sum_{a,b}^{a\leq b} \sum_{c,d}^{c\leq d} \bar{W}_{\ell}^{ab}\bar{W}_{\ell}^{cd}C_{\ell}^{a(c|}C_{\ell}^{b|d)}\right. \nonumber \\
		\phantom{asdas}\left.- \lambda \left(\sum^{c\leq d}_{c,d}\bar{W}_{\ell}^{cd}a^{c}a^{d}-1\right) \right] =0 \; \; ,i\leq j\\
		\sum^{c\leq d}_{c,d}\bar{W}_{\ell}^{cd} a^{c} a^{d} = 1
	\end{cases} \nonumber \\
	\implies &\begin{cases}
		\sum_{c,d}^{c\leq d} \bar{W}_{\ell}^{cd}C_{\ell}^{(c|i}C_{\ell}^{|d)j} + \bar{\lambda} a^{i}a^{j} =0 & \mbox{, $i\leq j$}\\
			\sum^{c\leq d}_{c,d}\bar{W}_{\ell}^{cd} a^{c} a^{d} = 1
	\end{cases} \; ,
	  \label{eq:pilc_eqs_2}
\end{align}
where $\bar{\lambda}\equiv -\lambda/2$. In matrix form,
\begin{widetext}
\begin{align}
	\begin{pmatrix}
		\Rxx^{(1|1}\Rxx^{|1)1} & \Rxx^{(1|1}\Rxx^{|2)1}  & \cdots & \Rxx^{(1|1}\Rxx^{|N)1} & \Rxx^{(2|1}\Rxx^{|2)1} & \Rxx^{(2|1}\Rxx^{|3)1} & \cdots & \Rxx^{(N|1}\Rxx^{|N)1} & a^1 a^1\\
		\Rxx^{(1|1}\Rxx^{|1)2} & \Rxx^{(1|1}\Rxx^{|2)2}  & \cdots & \Rxx^{(1|1}\Rxx^{|N)2} & \Rxx^{(2|1}\Rxx^{|2)2} &  \Rxx^{(2|1}\Rxx^{|3)2} & \cdots & \Rxx^{(N|1}\Rxx^{|N)2} & a^1 a^2 \\
		\vdots &\vdots&\ddots&\vdots&\vdots&\vdots&\ddots&\vdots& \vdots\\
		\Rxx^{(1|1}\Rxx^{|1)N} & \Rxx^{(1|1}\Rxx^{|2)N} & \cdots & \Rxx^{(1|1}\Rxx^{|N)N} & \Rxx^{(2|1}\Rxx^{|2)N} & \Rxx^{(2|1}\Rxx^{|3)N} & \cdots & \Rxx^{(N|1}\Rxx^{|N)N} & a^1 a^N \\
		\Rxx^{(1|2}\Rxx^{|1)2} & \Rxx^{(1|2}\Rxx^{|2)2}  & \cdots & \Rxx^{(1|2}\Rxx^{|N)2} & \Rxx^{(2|2}\Rxx^{|2)2} & \Rxx^{(2|2}\Rxx^{|3)2} & \cdots & \Rxx^{(N|2}\Rxx^{|N)2} & a^2 a^2\\
		\Rxx^{(1|2}\Rxx^{|1)3} & \Rxx^{(1|2}\Rxx^{|2)3}  & \cdots & \Rxx^{(1|2}\Rxx^{|N)3} & \Rxx^{(2|2}\Rxx^{|2)3} & \Rxx^{(2|2}\Rxx^{|3)3} & \cdots & \Rxx^{(N|2}\Rxx^{|N)3} & a^2 a^3\\
		\vdots &\vdots&\ddots&\vdots&\vdots&\vdots&\ddots&\vdots&\vdots\\
		\Rxx^{(1|N}\Rxx^{|1)N} & \Rxx^{(1|N}\Rxx^{|2)N} & \cdots & \Rxx^{(1|N}\Rxx^{|N)N} & \Rxx^{(2|N}\Rxx^{|2)N} & \Rxx^{(2|N}\Rxx^{|3)N} & \cdots & \Rxx^{(N|N}\Rxx^{|N)N} & a^N a^N\\
        a^1 a^1 & a^1 a^2 & \cdots & a^1 a^N & a^2 a^2 & a^2 a^3 & \cdots & a^N a^N & 0
	\end{pmatrix}
	\begin{pmatrix}
		\bar{W}_\ell^{11}\\
		\bar{W}_\ell^{12}\\
		\vdots\\
    	\bar{W}_\ell^{1N}\\
		\bar{W}_\ell^{22}\\
		\bar{W}_\ell^{23}\\
		\vdots\\
		\bar{W}_\ell^{NN}\\
		\bar{\lambda}
	\end{pmatrix} = 
	\begin{pmatrix}
		0\\0\\\vdots\\0\\0\\0\\\vdots\\0\\1
	\end{pmatrix} \; ,
    \label{eq:app_matrix_eq}
\end{align}
\end{widetext}
\begin{widetext}
We can further simply by defining the following vectors and matrices:
\begin{align}
	D_\ell \equiv 
	\underbrace{
	\begin{pmatrix}
		\Rxx^{(1|1}\Rxx^{|1)1} & \Rxx^{(1|1}\Rxx^{|2)1}  & \cdots & \Rxx^{(1|1}\Rxx^{|N)1} & \Rxx^{(2|1}\Rxx^{|2)1} & \Rxx^{(2|1}\Rxx^{|3)1} & \cdots & \Rxx^{(N|1}\Rxx^{|N)1}\\
		\Rxx^{(1|1}\Rxx^{|1)2} & \Rxx^{(1|1}\Rxx^{|2)2}  & \cdots & \Rxx^{(1|1}\Rxx^{|N)2} & \Rxx^{(2|1}\Rxx^{|2)2} &  \Rxx^{(2|1}\Rxx^{|3)2} & \cdots & \Rxx^{(N|1}\Rxx^{|N)2} \\
		\vdots &\vdots&\ddots&\vdots&\vdots&\vdots&\ddots&\vdots\\
		\Rxx^{(1|1}\Rxx^{|1)N} & \Rxx^{(1|1}\Rxx^{|2)N} & \cdots & \Rxx^{(1|1}\Rxx^{|N)N} & \Rxx^{(2|1}\Rxx^{|2)N} & \Rxx^{(2|1}\Rxx^{|3)N} & \cdots & \Rxx^{(N|1}\Rxx^{|N)N} \\
		\Rxx^{(1|2}\Rxx^{|1)2} & \Rxx^{(1|2}\Rxx^{|2)2}  & \cdots & \Rxx^{(1|2}\Rxx^{|N)2} & \Rxx^{(2|2}\Rxx^{|2)2} & \Rxx^{(2|2}\Rxx^{|3)2} & \cdots & \Rxx^{(N|2}\Rxx^{|N)2} \\
		\Rxx^{(1|2}\Rxx^{|1)3} & \Rxx^{(1|2}\Rxx^{|2)3}  & \cdots & \Rxx^{(1|2}\Rxx^{|N)3} & \Rxx^{(2|2}\Rxx^{|2)3} & \Rxx^{(2|2}\Rxx^{|3)3} & \cdots & \Rxx^{(N|2}\Rxx^{|N)3}\\
		\vdots &\vdots&\ddots&\vdots&\vdots&\vdots&\ddots&\vdots\\
		\Rxx^{(1|N}\Rxx^{|1)N} & \Rxx^{(1|N}\Rxx^{|2)N} & \cdots & \Rxx^{(1|N}\Rxx^{|N)N} & \Rxx^{(2|N}\Rxx^{|2)N} & \Rxx^{(2|N}\Rxx^{|3)N} & \cdots & \Rxx^{(N|N}\Rxx^{|N)N} \\
	\end{pmatrix}
}_{\text{$N(N+1)/2$ rows $\times N(N+1)/2$ columns}} \; . \label{eq:pilc_def_D}
\end{align}

\begin{align}
	\mathbf{t} &\equiv 
	\underbrace{\begin{pmatrix}
		a_1 a_1 & a_1 a_2 & \cdots & a_1 a_N & a_2 a_2 & a_2 a_3 \cdots a_N a_N 
\end{pmatrix}^{\text{T}}}_{\text{$N(N+1)/2$ rows}} \; ,\label{eq:pilc_def_t_app}\\
	\mathbf{w}_\ell &\equiv 
	\underbrace{\begin{pmatrix}
			\bar{W}_\ell^{11} & \bar{W}_\ell^{12} & \cdots & \bar{W}_\ell^{1N} & \bar{W}_\ell^{22} & \bar{W}_\ell^{23} \cdots \bar{W}_\ell^{NN}
	\end{pmatrix}^{\text{T}}}_{\text{$N(N+1)/2$ rows}} \; .
\label{eq:pilc_def_w_app}
\end{align}
\end{widetext}
We show that matrix $D_\ell^{\mu\nu}$ is symmetric, where Greek index ranges from 1 to $M$: Consider indices $\mu$ and $\nu$ such that $w_{\ell}^\mu = \bar{W}_\ell^{ab}$ and  $w_{\ell}^\nu = \bar{W}_\ell^{cd}$,
\begin{align}
	D_\ell^{\mu\nu} = \Rxx^{(a|c}\Rxx^{|b)d} &= \frac{1}{2}\left(\Rxx^{ac}\Rxx^{bd}+\Rxx^{bc}\Rxx^{ad} \right) \nonumber \\
			       &= \frac{1}{2}\left(\Rxx^{ca}\Rxx^{db} + \Rxx^{da}\Rxx^{cb} \right) \nonumber \\
			       &= \Rxx^{(c|a}\Rxx^{|d)b}  = D_\ell^{\nu\mu}\; .
\end{align}

The matrix equation Eq.\ (\ref{eq:app_matrix_eq}) simplifies to
\begin{align}
	\begin{pmatrix}
		D_{\ell}^{11} & D_{\ell}^{12} & \cdots & D_{\ell}^{1M} & t^1\\
		D_{\ell}^{21} & D_{\ell}^{22} & \cdots & D_{\ell}^{2M} & t^2 \\
		\vdots & \vdots & \ddots & \vdots & \vdots\\
		D_{\ell}^{M1} & D_{\ell}^{M2} & \cdots & D_{\ell}^{MM} & t^M\\
		t^1 & t^2 &\cdots & t^M & 0
	\end{pmatrix}
	\begin{pmatrix}
		w_{\ell}^1 \\ w_{\ell}^2 \\ \vdots \\ w_{\ell}^M \\ \bar{\lambda}
	\end{pmatrix}=
	\begin{pmatrix}
		0 \\ 0 \\ \vdots \\ 0 \\ 1
	\end{pmatrix} \; ,\label{eq:pilc_eqs_matrix}
\end{align}
where $M\equiv N(N+1)/2$. One recognizes this as the same matrix inversion problem for the original ILC, where the weight vector $\mathbf{w}_\ell$ in Eq.\ (\ref{eq:pilc_def_w_app}) is given by 
\begin{align}
	w_\ell^\mu = \frac{D^{-1}_{\mu\nu,\ell}t_\nu}{t_{\rho}D^{-1}_{\rho\tau,\ell}t_\tau} \; .
	\label{eq:pilc_weights_app}
\end{align}

\subsection{Derivation for Constrained SpILC weights}
\label{app:derivation_cspilc}
 Consider the two component model
\begin{align}
    x_p^i = a^i s_p + b^i y_p + n^i_p \; ,
\end{align}
Similar to Eq.\ (\ref{eq:pilc_eqs_2}), the \texttt{weak}cSpILC set of equations can be simplified to
\begin{align}
	&\begin{cases}
		\sum_{c,d}^N W_{\ell}^{cd} C_{\ell}^{ci} C_{\ell}^{dj} + \bar\lambda a^i a^j + \bar{\mu}b^i b^j =0 & \mbox{for $i\leq j$}\\
		W_{\ell}^{cd}a^c a^d = 1 \\
	 W_{\ell}^{cd} b^c b^d = 0 
	\end{cases} \nonumber \; ,  \\
	\implies &\begin{cases}
		\sum_{c,d}^{c\leq d} \bar{W}_\ell^{cd}C_{\ell}^{(c|i}C_{\ell}^{|d)j} + \bar{\lambda} a^{i}a^{j} + \bar\mu b^i b^j =0 & \mbox{, $i\leq j$}\\
			\sum^{c\leq d}_{c,d}\bar{W}_\ell^{cd} b^{c} b^{d} = 0\\
			\sum^{c\leq d}_{c,d}\bar{W}_\ell^{cd} a^{c} a^{d} = 1\\
	\end{cases} \; .
	  \label{eq:cpilc_eqs_2}
\end{align}
Using the definitions in Eqs.\ (\ref{eq:pilc_def_D})--(\ref{eq:pilc_def_w_app}), and
\begin{align}
	\mathbf{u}\equiv 
	\underbrace{\begin{pmatrix}
			 b^1b^1 & b^1b^2 & b^1 b^3 & \cdots & b^2b^2 & b^2 b^3 & \cdots & b^Nb^N
	\end{pmatrix}^{\text{T}}}_{\text{$N(N+1)/2$ rows}} \; .
\end{align}
The set of equations are cast into matrix form:
\begin{align}
	\begin{pmatrix}
		D_{\ell}^{11} & D_{\ell}^{12} & \cdots & D_{\ell}^{1M} & u^1 & t^1\\
		D_{\ell}^{21} & D_{\ell}^{22} & \cdots & D_{\ell}^{2M} & u^2 & t^2 \\
		\vdots & \vdots & \ddots & \vdots &\vdots & \vdots\\
		D_{\ell}^{M1} & D_{\ell}^{M2} & \cdots & D_{\ell}^{MM} & u^M& t^M\\
		u^1 & u^2 &\cdots & u^M & 0 & 0\\
		t^1 & t^2 &\cdots & t^M & 0 & 0
	\end{pmatrix}
	\begin{pmatrix}
		w_\ell^1 \\ w_\ell^2 \\ \vdots \\ w_\ell^M \\ \bar{\mu} \\ \bar\lambda
	\end{pmatrix}=
	\begin{pmatrix}
		0 \\ 0 \\ \vdots \\ 0 \\ 0 \\ 1
	\end{pmatrix} \; .\label{eq:pilc_eqs_matrix}
\end{align}

Analogous to the derivation of constrained ILC weights, the \texttt{weak}cSpILC weights are
 \begin{align}
	 w_\ell^\mu &= \frac{\left( \mathbf{u}^\text{T}\mathbf{D}_\ell^{-1}\mathbf{u} \right) D^{-1}_{\ell,\mu\rho} t^\rho-\left(\mathbf{t}^\text{T}\mathbf{D}_\ell^{-1}\mathbf{u}\right) D^{-1}_{\ell,\mu\rho}u^\rho}{\left(\mathbf{t}^\text{T}\mathbf{D}_\ell^{-1}\mathbf{t}\right)\left( \mathbf{u}^\text{T}\mathbf{D}_\ell^{-1}\mathbf{u} \right)- \left(\mathbf{t}^\text{T}\mathbf{D}_\ell^{-1}\mathbf{u}\right)^2} \; .
	 \label{eq:cspilc_weights_app}
\end{align}
\subsection{Non-Gaussian contributions to the variance}
\label{sec:gaussianity}
We have not assumed component Gaussianity when we derive SpILC weights. We make this argument by showing that the variance does not depend on higher-point statistics of the modeled components. Consider the two component model
\begin{align}
    x_p^i = a^i s_p + b^i y_p + n^i_p \; .
\end{align}
For \texttt{strong}-cSpILC (equivalent to cILC), we impose the constraints $W^{ij}_\ell a^i a^j = 1$, $W^{ij}_\ell b^i b^j = 0$ and $W^{ij}_\ell a^{(i}b^{j)}=0$.
The variance of $\hat{K}_\ell$ can be expanded in terms of four-point and two-point functions. Schematically, the four-point functions are $\langle ssss \rangle$, $\langle sssy \rangle$, $\langle sssn \rangle$, $\langle ssyn \rangle$, $\langle ssnn \rangle$, $\langle sysy \rangle$, $\langle sysn \rangle$, $\langle syyn \rangle$, $\langle synn \rangle$, $\langle snsn \rangle$, $\langle snyn \rangle$, $\langle snnn \rangle$, $\langle ynyn \rangle$, $\langle ynnn\rangle$, and $\langle nnnn \rangle$. Applying the deprojection constraints, the remaining non-vanishing four-point functions are $\langle ssss\rangle$ and those involving some factors of $n$. Since $n$ is independent of $s$ and $y$, and is assumed to be Gaussian and zero-mean, the non-vanishing 4-point functions are $\langle ssss\rangle$, $\langle ssnn \rangle$, $\langle snsn\rangle$, $\langle snyn\rangle$, $\langle ynyn\rangle$ and $\langle nnnn\rangle$. 
Noting that e.g.\ $s$ and $n$ being uncorrelated implies
\begin{align}
\langle ssnn \rangle = \langle ss \rangle \langle nn \rangle \; ,
\end{align}
the only four-point function that remains is $\langle ssss\rangle$ and $\langle nnnn\rangle$. Performing the variance minimization
\begin{align}
	\frac{\partial}{\partial W_{\ell}^{jk}}\operatorname{Var}(\hat{K}_\ell) &= \frac{\partial}{\partial W_{\ell}^{jk}} \left[ \operatorname{Var}(\hat{C}_\ell^{ss}) +\cdots \right] \; ,
\end{align}
the only four-point function in the variance that involves the components, $\langle ssss\rangle$, is independent of $W^{jk}_\ell$, so its derivative is zero. The implication is as follows: \emph{the SpILC weights are optimal even for non-Gaussian components}, but calculations of $\operatorname{Var}(\hat{K}_\ell)$ using Eq.\ (\ref{eq:gaussian_variance_estimate}) will be incorrect if the connected part of $\langle ssss\rangle$ is non-zero. Note however that the optimality of the estimator does depend on the independence and Gaussianity of \emph{noise}, as the Gaussian approximation $\operatorname{Cov}(\hat{N}^{jk},\hat{N}^{mn})= 2\hat{N}^{jm}\hat{N}^{kn}/N_p$ is used.

For \texttt{weak}-cSpILC, we relax the assumption of $W^{ij}_\ell a^{(i}b^{j)}=0$. Applying the deprojection constraints, the remaining non-vanishing four-point functions are $\langle ssss\rangle$, $\langle sssy \rangle$, $\langle sysy \rangle$, and four-point functions involving $n$ which reduces to two-point functions as above, except for $\langle nnnn \rangle$. Therefore the only non-Gaussian contribution to the variance is:
\begin{align}
    \langle s_p s^*_p s_q y^*_q \rangle_c = \langle s_p y^*_p s_q y^*_q \rangle_c = 0 \; .
\end{align}
For $s_p$ being the lensed CMB+kSZ, and $y_p$ a parity-even component, e.g.\ tSZ, the kSZ$\times$tSZ$\times$kSZ$\times$tSZ connected four-point function is non-vanishing.

\section{Equivalences in estimator weights}
\subsection{Equivalence between SpILC and ILC}
\label{app:eqv_ilc_pilc}
We provide a short argument here to explain why we have such an equality between standard ILC and SpILC spectra. Suppose $w^{i}_{\ell,\text{ILC}}$ solves the ILC constrained-minimization problem, such that $w_{i}^{\text{ILC}}$ and the Lagrange multiplier $\lambda_\text{ILC}\equiv -2\bar\lambda_\text{ILC}$ satisfies 
\begin{align}
w^{j}_{\ell,\text{ILC}} \hat{C}^{ji}_\ell  + \bar{\lambda}_\text{ILC} a^i &=0 \; ,\label{eq:w_ilc_prop1}\\
	w^{j}_{\ell,\text{ILC}} a^j &= 1 \; . \label{eq:w_ilc_prop2}
\end{align}
We can manipulate Eq.\ (\ref{eq:w_ilc_prop1}) to obtain
\begin{align}
	&(w^{m}_{\ell,\text{ILC}}\hat{C}_\ell^{mi}+\bar{\lambda}_\text{ILC}a^i)(w^{n}_{\ell,\text{ILC}}\hat{C}_\ell^{nj}-\bar{\lambda}_\text{ILC}a^j) = 0 \nonumber \\
	\implies& w^{m}_{\ell,\text{ILC}}w^{n}_{\ell,\text{ILC}}\hat{C}_\ell^{mi}\hat{C}_\ell^{nj} - \bar\lambda_\text{ILC}^2 a^i a^j =2 \bar\lambda_\text{ILC} a^{[j|} w^{m}_{\ell,\text{ILC}} \hat{C}_\ell^{m|i]} \nonumber\\
	\implies& w^{m}_{\ell,\text{ILC}}w^{n}_{\ell,\text{ILC}}\hat{C}_\ell^{mi}\hat{C}_\ell^{nj} - \bar\lambda_\text{ILC}^2 a^i a^j =-2 \bar\lambda_\text{ILC}^2 a^{[j} a^{i]}=0 \;,
\label{eq:w_ilc_prop3}
\end{align}
where the anti-symmetrization bracket is defined with $a^{[i}b^{j]}\equiv (a^i b^j - a^j b^i)/2$. We made use of the property $w^{j}_{\ell,\text{ILC}} \hat{C}^{ji}_\ell = -\bar\lambda_\text{ILC} a^i$ to arrive at the third equality. Taking the square of Eq.\ (\ref{eq:w_ilc_prop2}),
\begin{align}
	w_{\ell,\text{ILC}}^i w_{\ell,\text{ILC}}^j a^i a^j &= 1 \; .
	\label{eq:w_ilc_prop4}
\end{align}
From Eqs.\ (\ref{eq:sum_manipulation}) and (\ref{eq:pilc_eqs_2}), the standard SpILC simultaneous equations can be written as
\begin{align}
	\begin{cases}
		\sum_{c,d}^N W^{cd}_\ell \hat{C}_\ell^{ci} \hat{C}_\ell^{dj} + \bar\lambda a^i a^j =0 & \mbox{for $i\leq j$}\\
		W^{cd}_\ell a^c a^d = 1
	\end{cases} \; .
	\label{eq:pilc_eq_simplified}
\end{align}
From Eqs.\ (\ref{eq:w_ilc_prop3}) and (\ref{eq:w_ilc_prop4}), one finds that $W_\ell^{ij} = w_{\ell,\text{ILC}}^i w_{\ell,\text{ILC}}^j$ and  $\bar\lambda = -\bar\lambda^2_\text{ILC}$ is a solution to the standard SpILC set of equations. Because the solution to Eq.\ (\ref{eq:pilc_eqs_matrix}) is unique, 
\begin{align}
	W_{\ell,\text{SpILC}}^{ij} = w_{\ell,\text{ILC}}^i w_{\ell,\text{ILC}}^j \; .
\end{align}
We can also show that standard SpILC (equivalently, standard ILC) is equivalent to the power spectrum estimator in Sec.\ 3.6 of Ref.\ \cite{tegmark_efstathiou}, which seeks an internal linear combination of frequency maps that minimizes the bias of the power spectrum of an independent component (uncorrelated with foregrounds and noise). Writing their optimization problem as
\begin{align}
	\begin{cases}
		w^i_\ell a^i = 1 \; ,\\
		\partial_{w^i_\ell} (\hat{C}^{ss}_\ell-C^{ss}_\ell) = \partial_{w^i_\ell} \hat{C}^{ss}_\ell = \partial_{w_i} w_j w_k \hat{C}^{jk}_\ell = 0 \; ,
	\end{cases}
\end{align}
the optimization problem and weights are equivalent to that of standard harmonic ILC. As we proved above, standard ILC is equivalent to standard SpILC in the spectral level in the presence of Gaussian noise, in other words minimizing the bias of the power spectrum also minimizes its (Gaussian) variance.

\subsection{Equivalence between \texttt{strong}-cSpILC and cILC}
\label{app:eqv_cilc_cpilc}

Consider the three-component data model
\begin{align}
	x^i_p &= a^i s_p + b^i y_p + c^i z_p + n^i_p \; ,
\end{align}
where the power of the signal $s_p$ is to be recovered. Constrained ILC gives the following set of equations
\begin{align}
	&\begin{cases}
		w_{\ell}^j C_{\ell}^{ji} + \bar{\lambda} a^i + \bar{\mu} b^i +\bar{\nu} c^i =0 \\
		w_{\ell}^i a^i = 1 \\
		w_{\ell}^i b^i = 0 \\
		w_{\ell}^i c^i = 0 \; .
	\end{cases}
	\label{eq:app_cilc_equation}
\end{align}
where $\lambda, \mu,\nu$ are undetermined Lagrange multipliers,  $\bar\lambda\equiv -\lambda/2$ and similarly for  $\bar\mu$ and  $\bar\nu$.

Now turn to the \texttt{strong}SpILC estimator, which we show to be equivalent to the cILC system:
\begin{align}
	&\begin{cases}
		\frac{\partial}{\partial W_{\ell}^{ij}}\left[W_{\ell}^{ab}W_{\ell}^{cd}C_{\ell}^{ac}C_{\ell}^{bd} -\lambda' (W_{\ell}^{cd}a^c a^d-1)\right. \nonumber \\
		\phantom{AAAAA}\left.-\mu' W_{\ell}^{cd}b^c b^d - \nu' W_{\ell}^{cd}c^c c^d -2\alpha' W_{\ell}^{cd} a^{(c}b^{d)}  \right.\nonumber \\
		\phantom{AAAAAAAAA}\left.-2\beta' W_{\ell}^{cd} a^{(c} c^{d)} - 2\gamma' W_{\ell}^{cd}b^{(c}c^{d)}\right] =0 \\
		W_{\ell}^{ij} a^i a^j = 1 \\
		W_{\ell}^{ij} b^i b^j =0 \\
		W_{\ell}^{ij} c^i c^j =0 \\
		W_{\ell}^{ij} a^{(i} b^{j)} =0 \\
		W_{\ell}^{ij} a^{(i} c^{j)} =0 \\
		W_{\ell}^{ij} b^{(i} c^{j)} =0 \\
	\end{cases}\\ \implies 
	&\begin{cases}
		W_{\ell}^{cd}C_{\ell}^{ic}C_{\ell}^{jd} +\bar\lambda' a^i a^j +\bar\mu' b^i b^j +\bar\nu' c^i c^j \\
		\phantom{AAAAA}+2\bar\alpha'a^{(i}b^{j)} +2\bar\beta'  a^{(i} c^{j)} + 2\bar\gamma' b^{(i}c^{j)} =0 \\
		W_{\ell}^{ij} a^i a^j = 1 \\
		W_{\ell}^{ij} b^i b^j =0 \\
		W_{\ell}^{ij} c^i c^j =0 \\
		W_{\ell}^{ij} a^{(i} b^{j)} =0 \\
		W_{\ell}^{ij} a^{(i} c^{j)} =0 \\
		W_{\ell}^{ij} b^{(i} c^{j)} =0 \\
	\end{cases}\label{eq:app_cPILC} \; ,
\end{align}
where $\lambda',\mu',\nu',\alpha',\beta',\gamma'$ are Lagrange multipliers,  $\bar\lambda'\equiv -\lambda'/2$ and similarly for the rest.
We wish to show that the weights $w^i_\ell$ of the cILC system also solves the set of \texttt{strong}cSpILC equations, i.e.\
 \begin{align}
	 W^{cd}_{\ell,\mathtt{strong}\text{cSpILC}} = w^c_{\ell,\text{cILC}} w^d_{\ell,\text{cILC}} \; .
\end{align}
Starting from the cILC equations,
\begin{align}
	&w_{\ell}^c C_{\ell}^{ci} + \bar\lambda a^i + \bar\mu b^i + \bar\nu c^i =0 \nonumber \\
	\implies &(w_{\ell}^c C_{\ell}^{ci} + \bar\lambda a^i + \bar\mu b^i + \bar\nu c^i)(w_{\ell}^d C_{\ell}^{dj}-\bar\lambda a^j -\bar\mu b^j -\bar\nu c^j)\nonumber \\
		 &\hspace{45ex}=0 \nonumber \\
	\implies & w_{\ell}^c w_{\ell}^d C_{\ell}^{ci}C_{\ell}^{dj} - (\bar\lambda a^i + \bar\mu b^i + \bar\nu c^i)(\bar\lambda b^j + \bar\mu b^j + \bar\nu c^j) \nonumber \\
		 &- w_{\ell}^c C_{\ell}^{ci}(\bar\lambda a^j + \bar\mu b^j + \bar\nu c^j) + w_{\ell}^d C_{\ell}^{dj}(\bar\lambda a^i +\bar\mu b^i + \bar\nu c^i)\nonumber \\
		 &\hspace{45ex}=0 \; .
		 \label{eq:app_cilc_manipulation}
\end{align}
We can cancel the final two terms on the LHS of Eq.\ (\ref{eq:app_cilc_manipulation}) using Eq.\ (\ref{eq:app_cilc_equation}),
\begin{align}
	w_{\ell}^c C_{\ell}^{ci} = -(\bar\lambda a^i + \bar\mu b^i + \bar\nu c^i) \; ,
\end{align}
giving
\begin{align}
	 & 
	 w_{\ell}^c w_{\ell}^d C_{\ell}^{ci}C_{\ell}^{dj} - (\bar\lambda a^i + \bar\mu b^i + \bar\nu c^i)(\bar\lambda b^j + \bar\mu b^j + \bar\nu c^j) \nonumber \\
		 & +w_{\ell}^c C_{\ell}^{ci}w_{\ell}^d C_{\ell}^{dj} - w_{\ell}^d C_{\ell}^{dj} w_{\ell}^c C_{\ell}^{ci} =0 \nonumber \\
	\implies &w_{\ell}^c w_{\ell}^d C_{\ell}^{ci}C_{\ell}^{dj} - (\bar\lambda a^i + \bar\mu b^i + \bar\nu c^i)(\bar\lambda b^j + \bar\mu b^j + \bar\nu c^j) =0 \; .
\end{align}
Therefore, $(W^{cd}_\ell,\bar\lambda',\bar\mu',\bar\nu',\bar\alpha',\bar\beta',\bar\gamma')$ equals to $(w_{\ell}^c w_{\ell}^d,-\bar\lambda^2,-\bar\mu^2,-\bar\nu^2,-\bar\lambda\bar\mu,-\bar\lambda\bar\nu,-\bar\mu\bar\nu)$ is a solution to the \texttt{strong}cSpILC equations Eq.\ (\ref{eq:app_cPILC}). Because the solution is unique, this verifies that the ILC-SpILC equivalence generalizes to multiple components.

\section{Incorporating Data Splits}
\subsection{Data-split ILC and cILC}

Suppose $x^i_p$ is a map constructed from data collected from time $t_0$ to  $t_0 + \Delta t$. Two noisier maps $x^{i,\ds}_p$ and $x^{i,\dss}_p$, referred to as \emph{data splits}, can be constructed from the time segments $[t_0,t_0+\Delta t/2]$ and  $[t_0+\Delta t/2,t_0+\Delta t]$ respectively.  The two time segments are labeled 1 and 2 respectively. The noise in different data splits are assumed to be uncorrelated.

An ILC method built on these splits was proposed in Ref.\ \cite{hill_spergel}, where the original motivation is to minimize the foreground variance in the auto-spectra of ILC and cILC maps by preventing instrumental noise from contributing to the weights. Here we review this approach, show that it is not the minimum-variance power spectrum estimator, and derive the minimum-variance data-split estimator using the SpILC method proposed in this work.

Ref.\ \cite{hill_spergel} proposes to produce ILC (cILC) maps $\hat{s}^{\ds}_p \equiv w^i_\ell x^{i,\ds}_p$ and  $\hat{s}^{\dss}_p \equiv w^i_\ell x^{i,\dss}_p$ where weights are determined by minimizing the sample covariance between two different splits
\begin{align}
	\hat{\sigma}^2_{\ds\dss}\equiv \frac{1}{N_p}\sum_p \hat{s}^{\ds}_p \hat{s}^{*,\dss}_p \; ,
\end{align}
subject to ILC (or cILC) constraints. The auto-spectrum of the signal is then estimated by $\hat{\sigma}^2_{\ds\dss}$. 
The elimination of the noise bias can be simply demonstrated:
\begin{align}
	\langle \hat{\sigma}^2_{\ds\dss}  \rangle &= \frac{1}{N_p}\sum_p \langle (s_p + w^i_\ell k^i_{p}+ w_{\ell}^i n^i_{p,\ds})(s^*_p +w^j_\ell k^{*,j}_p + w_{\ell}^j n^{*,j}_{p,\dss})  \rangle \nonumber \\
						  &= C^{ss}_\ell + w^i_\ell w^j_\ell \frac{1}{N_p}\sum_p \langle k^i_p k^{*,j}_p  \rangle  + w_{\ell}^i w_{\ell}^j\frac{1}{N_p}\sum_p \langle n^i_{p,\ds} n^{*,j}_{p,\dss}  \rangle \nonumber \\
						  &= C^{ss}_\ell + w^i_\ell w^j_\ell F^{ij}_\ell  \; , \label{app:split_expectation}
\end{align}
where $k^i_p$ is the foreground map at the  $i$-th frequency, and  $F^{ij}_\ell\equiv \langle k^i_p k^j_p  \rangle$ is the foreground cross spectra. Here we assumed that foregrounds are independent from the CMB and instrumental noise. If cILC constraints are imposed, then the foreground term $w^i_\ell w^j_\ell F^{ij}_\ell$ can be further reduced or eliminated at a noise cost.

\subsection{Optimal weights for the data-split ILC and cILC power spectrum estimators}
Constructing data splits that use weights that minimizes $\hat{\sigma}^2_{12}$ minimizes the foreground sample covariance $\sum_p w^j_\ell w^k_\ell k_p^{j}k_p^{k*} /N_p$ together with sample covariance due to chance correlations between signal, foreground and noise in the two splits. 
If cILC constraints are imposed, even the foreground sample covariance vanishes, and the weights only minimizes these chance correlations i.e.\ ILC biases.

Instead, if one wishes to minimize the variance of the estimated auto power spectrum $\hat{\sigma}^2_{12}$, the weights discussed above is not optimal. Writing out the variance:
\begin{align}
	\operatorname{Var}(\hat{\sigma}^2_{\ds\dss}) &= \frac{2}{N_p} w_{\ell}^j w_{\ell}^k w_{\ell}^m w_{\ell}^n C^{jm}_{\ell,\ds(\ds|}C^{kn}_{\ell,\dss|\dss)} \nonumber \\
						      &= \frac{1}{N_p} \left( C^{ss}_\ell + w^j_\ell w^k_\ell F^{jk}_\ell\right)^2 \nonumber \\
						      &\phantom{{}={}} +\frac{1}{N_p} \left( C^{ss}_\ell + w^j_\ell w^k_\ell F^{jk}_\ell+  2w_{\ell}^j w_{\ell}^k N_\ell^{jk}\right)^2 \; ,
\end{align}
where we made the simplifying assumptions of Gaussianity, and temporal stationarity of data such that
\begin{align}
	\langle n^j_{p,\ds} n^k_{p,\ds}  \rangle = \langle n^j_{p,\dss} n^k_{p,\dss}  \rangle = 2\langle n^j_p n^k_p  \rangle \equiv 2 N_\ell^{jk} \; .
\end{align}
The weights above which neglects the noise power $w^i_\ell w^j_\ell N^{ij}_\ell$ does not minimize the variance of the signal auto-spectrum estimator, $\operatorname{Var}(\hat{\sigma}^2_{\ds\dss})$. 

On the other hand, note that the normal ILC and cILC weights (using the ensemble covariance) minimize (subject to constraints)
\begin{align}
	\operatorname{Var}(\hat{s}^\text{ILC}) &=
	C^{ss}_\ell+w^j_\ell w^k_\ell(F^{jk}_\ell + N^{jk}_\ell) \; , \\
	\operatorname{Var}(\hat{s}^\text{cILC}) &=
	C^{ss}_\ell+w^j_\ell w^k_\ell N^{jk}_\ell  \; .
\end{align}
One therefore sees that the cILC weights are the optimal weights for minimizing the variance $\operatorname{Var}(\hat{\sigma}^2_{12})$ of the cILC data-split power spectrum. However, neither the ILC nor the ILC data-split weights above minimizes $\operatorname{Var}(\hat{\sigma}^2_{12})$. To build a data-split power spectrum estimator with minimum variance one has to use the SpILC formalism, discussed in the next section.
\onecolumngrid
\subsection{Variance of Data-Split Estimators}
\label{app:splits_c}
Depending on the observational setup, the noise may be independent across different frequency channels or may have correlations across frequencies (which is the case for e.g. atmospheric noise). Here we propose two data-split power spectrum estimators: the first (DS0) uses only cross spectra between splits and is applicable when the noise is correlated across frequency channels, and the second (DS, considered in the main text) discards only the split auto spectra (same split and same frequency) and is applicable for the independent noise case.

	The first estimator generalizes the estimator in the previous sub-section:
\begin{align}
	\hat{K}_\ell^{\text{(DS0)}} = \sum_{ij}W^{ij}_\ell \hat{C}_{\ell,12}^{ij} \; ,
\end{align}
the variance is
\begin{align}
	\operatorname{Var}(\hat{K}_\ell^{\text{(DS0)}}) &= \frac{2}{2\ell+1} W^{jk}_\ell W^{mn}_\ell C^{jm}_{\ell,1(1|} C^{kn}_{\ell,2|2)} \; .
\end{align}
The expressions for the data-split weights are calculated through the replacement in the $D^{\mu\nu}_\ell$ matrix in Eq.\ (\ref{eq:pilc_def_D}) to (summation not implied for repeated indices, and color / subscript Greek $\alpha,\beta$ indices denote the set of indices to be symmetrized):
\begin{align}
	\hat{C}_\ell^{(m|i}\hat{C}_\ell^{|n)j} \mapsto& \hat{C}^{{\color{red}(m_\alpha|}{\color{blue}(i_\beta|}}_{\ell,(\ds|\ds}\hat{C}^{{\color{red}|n_\alpha)}{\color{blue}|j_\beta)}}_{\ell,|\dss)\dss} 
				\label{eq:split0_replacement} \; .
\end{align}
These symmetrization brackets are necessary to ensure that the object is symmetric in $j\leftrightarrow k$ and  $m\leftrightarrow n$ during the optimization. Once this is substituted in the  $D^{\mu\nu}_\ell$ matrix and Eq.\ (\ref{eq:pilc_eqs_matrix}) is solved, one gets the weights for $\hat{K}^{(\text{DS0})}_\ell$.
  
Now consider the second data-split estimator, which is considered in the main text in Eq.\ (\ref{eq:data_split1}), where only same-split, same-frequency information is discarded:
\begin{align}
	\hat{K}_\ell^\text{(DS)} &= \sum_{ij}^N W_{\ell}^{ij} \left[\delta_{ij}\hat{C}^{ij}_{\ell,\ds\dss} + (1-\delta_{ij})\hat{C}_\ell^{ij}\right] \; .
\end{align}
To build this estimator strictly with data-split maps, one replaces in $\hat{K}_\ell^{(\text{DS})}$
\begin{align}
	\hat{C}^{ij}_\ell  \mapsto&  \frac{1}{4N_p}\sum_p (x^i_{p,\ds} +x^i_{p,\dss})(x^{*,j}_{p,\ds}+x^{*,j}_{p,\dss})  \nonumber \\
							   &= \frac{1}{2} \left(\hat{C}^{ij}_{\ell,\ds(\ds|}+\hat{C}^{ij}_{\ell,\dss|\dss)}\right) \; .
\end{align}
The variance of this estimator is computed to be
\begin{align}
	\operatorname{Var}(\hat{K}_\ell^{(\mathrm{DS})}) &= \langle \hat{K}_\ell^{(\mathrm{DS})}\hat{K}_\ell^{(\mathrm{DS})}\rangle -\langle \hat{K}_\ell^{(\mathrm{DS})}\rangle^2\nonumber \\
						    &=\frac{1}{N^2_p}W_{\ell}^{jk}W_{\ell}^{mn} \sum_{p,q}\delta_{jk}\delta_{mn}\left[\langle x^j\p x^k\pp x^m\q x^n\qq  \rangle -\langle x^j\p x^k\pp  \rangle\langle  x^m\q x^n\qq  \rangle \right] + \nonumber \\
						    &\phantom{{}={}}\frac{1}{N^2_p}W_{\ell}^{jk}W_{\ell}^{mn} \sum_{p,q}\delta_{jk}(1-\delta_{mn})\left[\langle x^j\p x^k\pp x^m_q x^n_q  \rangle -\langle x^j\p x^k\pp  \rangle\langle  x^m_q x^n_q  \rangle \right] + \nonumber \\
						    &\phantom{{}={}}\frac{1}{N^2_p}W_{\ell}^{jk}W_{\ell}^{mn} \sum_{p,q}(1-\delta_{jk})\delta_{mn}\left[\langle x^j_p x^k_p x^m\q x^n\qq  \rangle -\langle x^j_p x^k_p  \rangle\langle  x^m\q x^n\qq  \rangle \right] + \nonumber \\
						    &\phantom{{}={}}\frac{1}{N^2_p}W_{\ell}^{jk}W_{\ell}^{mn} \sum_{p,q}(1-\delta_{jk})(1-\delta_{mn})\left[\langle x^j_p x^k_p x^m_q x^n_q  \rangle -\langle x^j_p x^k_p  \rangle\langle  x^m_q x^n_q  \rangle \right]\nonumber 
                            \end{align}
Applying Wick's theorem,
                            \begin{align}
						   \operatorname{Var}(\hat{K}_\ell^{(\mathrm{DS})})  &=\frac{1}{N^2_p}W_{\ell}^{jk}W_{\ell}^{mn} \sum_{p,q}\delta_{jk}\delta_{mn}\left[ \langle x^j\p x^m\q  \rangle \langle x^k\pp x^n\qq  \rangle + \langle x^j\p x^n\qq  \rangle\langle x^k\pp x^m\q  \rangle \right] + \nonumber \\
						    &\phantom{{}={}}\frac{1}{N^2_p}W_{\ell}^{jk}W_{\ell}^{mn} \sum_{p,q}\delta_{jk}(1-\delta_{mn})\left[ \langle x^j\p x^m_q  \rangle\langle x^k\pp x^n_q  \rangle + \langle  x^j\p x^n_q  \rangle\langle x^k\pp x^m_q  \rangle \right] + \nonumber \\
						    &\phantom{{}={}}\frac{1}{N^2_p}W_{\ell}^{jk}W_{\ell}^{mn} \sum_{p,q}(1-\delta_{jk})\delta_{mn}\left[\langle x^j_p x^m\q  \rangle \langle x^k_p x^n\qq  \rangle+ \langle x^j_p x^n\qq  \rangle \langle x^k_p x^m\q  \rangle \right] + \nonumber \\
						    &\phantom{{}={}}\frac{1}{N^2_p}W_{\ell}^{jk}W_{\ell}^{mn} \sum_{p,q}(1-\delta_{jk})(1-\delta_{mn})\left[ \langle  x^j_p x^m_q  \rangle \langle x^k_p x^n_q  \rangle + \langle  x^j_p x^n_q  \rangle\langle x^k_p x^m_q  \rangle\right]\nonumber \\
						    &=\frac{2}{N^2_p}W_{\ell}^{jk}W_{\ell}^{mn} \sum_{p,q}\left[\delta_{jk}\delta_{mn} \langle x^j\p x^m_{(q_1|}  \rangle \langle x^k\pp x^n_{|q_2)}  \rangle  + \delta_{jk}(1-\delta_{mn}) \langle x^j\p x^m_q  \rangle\langle x^k\pp x^n_q  \rangle \right] + \nonumber \\
						    &\phantom{{}={}}\frac{2}{N^2_p}W_{\ell}^{jk}W_{\ell}^{mn} \sum_{p,q}\left[(1-\delta_{jk})\delta_{mn}\langle x^j_p x^m\q  \rangle \langle x^k_p x^n\qq  \rangle +(1-\delta_{jk})(1-\delta_{mn}) \langle  x^j_p x^m_q  \rangle \langle x^k_p x^n_q  \rangle \right] \\
	\implies \operatorname{Var}(\hat{K}_\ell^{(\mathrm{DS})}) &= \frac{2}{2\ell+1}W_{\ell}^{jk}W_{\ell}^{mn}\left[\delta_{jk}\delta_{mn} {C}^{jm}_{\ell,\ds{(\ds|}} {C}^{kn}_{\ell,\dss{|\dss)}}+\frac{2}{4}\delta_{jk}(1-\delta_{mn} ) ({C}^{jm}_{\ell,\ds \ds}+{C}^{jm}_{\ell,\ds \dss}) ({C}^{kn}_{\ell,\dss \ds}+{C}^{kn}_{\ell,\dss \dss})	 \right. \nonumber\\
    &\phantom{{}={} \frac{2}{2\ell+1}W_{\ell}^{jk}W_{\ell}^{mn} }\left. \ +(1-\delta_{jk})(1-\delta_{mn} ) {C}^{jm}_\ell {C}^{kn}_\ell \right] \; .
\end{align}

The expressions for the data-split weights are calculated through the replacement in the $D^{\mu\nu}_\ell$ matrix in Eq.\ (\ref{eq:pilc_def_D}) to (summation not implied for repeated indices, and color / subscript Greek $\alpha,\beta$ indices denote the set of indices to be symmetrized):
\begin{align}
	\hat{C}_\ell^{(m|i}\hat{C}_\ell^{|n)j} \mapsto& \delta_{ij}\delta_{mn} \hat{C}^{{\color{red}(m_\alpha|}{\color{blue}(i_\beta|}}_{\ell,(\ds|\ds}\hat{C}^{{\color{red}|n_\alpha)}{\color{blue}|j_\beta)}}_{\ell,|\dss)\dss} \nonumber\\
    &+\delta_{ij} (1-\delta_{mn})\hat{C}^{{\color{red}(m_\alpha|}{\color{blue}(i_\beta|}}_{\ell,11+21}\hat{C}^{{\color{red}|n_\alpha)}{\color{blue}|j_\beta)}}_{\ell,12+22} \nonumber \\
					    &+(1-\delta_{ij})\delta_{mn}\hat{C}^{{\color{red}(m_\alpha|}{\color{blue}(i_\beta|}}_{\ell,11+12} \hat{C}^{{\color{red}|n_\alpha)}{\color{blue}|j_\beta)}}_{\ell,21+22}\nonumber \\  
					    &+(1-\delta_{ij})(1-\delta_{mn})\hat{C}_\ell^{(m|i}\hat{C}^{|n)j}_\ell\; ,
				\label{eq:split_replacement}
\end{align}
where
\begin{align}
    \hat{C}_{\ell,11+12}^{ij} \equiv \frac{1}{2} (\hat{C}^{ij}_{11}+\hat{C}^{ij}_{12}) \; ,
\end{align}
and similarly for $\hat{C}^{ij}_{\ell,21+22}$, $\hat{C}^{ij}_{\ell,11+21}$ and $\hat{C}^{ij}_{\ell,12+22}$.

\end{document}